\PassOptionsToPackage{prologue,dvipsnames,table}{xcolor}
\documentclass[sigconf]{acmart}

\copyrightyear{2026}
\acmYear{2026}
\setcopyright{cc}
\setcctype{by}
\acmConference[KDD 2026] {Proceedings of the 32nd ACM SIGKDD Conference on Knowledge Discovery and Data Mining V.1}{August 9--13, 2026}{Jeju Island, Republic of Korea.}
\acmBooktitle{Proceedings of the 32nd ACM SIGKDD Conference on Knowledge Discovery and Data Mining V.1 (KDD 2026), August 9--13, 2026, Jeju Island, Republic of Korea}
\acmISBN{979-8-4007-2258-5/2026/08}
\acmDOI{10.1145/3770854.3780264}
\usepackage{booktabs}
\usepackage{balance}
\usepackage{enumitem}
\usepackage{adjustbox}

\usepackage{rotating}
\usepackage{amsmath}
\usepackage{pifont}
\usepackage{array}
\usepackage{bbm}
\usepackage{multirow}
\usepackage{makecell}
\usepackage[normalem]{ulem}
\useunder{\uline}{\ul}{}
\usepackage{graphicx}
\usepackage{subcaption}
\usepackage[linesnumbered,algoruled,boxed,lined]{algorithm2e}
\usepackage{siunitx}

\newcommand{\cmark}{\ding{51}}
\newcommand{\xmark}{\ding{55}}

\newcommand{\ie}{{\it i.e.}}
\newcommand{\eg}{{\it e.g.}}

\newcolumntype{H}{>{\columncolor{orange!30}\color{black}}c}

\begin{document}
\title{MergeRec: Model Merging for Data-Isolated Cross-Domain Sequential Recommendation}

\author{Hyunsoo Kim} \authornotemark[1]
\affiliation{
  \institution{Sungkyunkwan University} \city{Suwon} \country{Republic of Korea}}
\email{khs1778@skku.edu}

\author{Jaewan Moon} \authornote{Both authors contributed equally to this research.}
\affiliation{
  \institution{Sungkyunkwan University} \city{Suwon} \country{Republic of Korea}}
\email{jaewan7599@skku.edu}

\author{Seongmin Park}
\affiliation{
  \institution{Sungkyunkwan University} \city{Suwon} \country{Republic of Korea}}
\email{psm1206@skku.edu}

\author{Jongwuk Lee}\authornote{Corresponding author}
\affiliation{
  \institution{Sungkyunkwan University} \city{Suwon} \country{Republic of Korea}}
\email{jongwuklee@skku.edu}



\begin{abstract}
Modern recommender systems trained on domain-specific data often struggle to generalize across multiple domains.
Cross-domain sequential recommendation has emerged as a promising research direction to address this challenge; however, existing approaches face fundamental limitations, such as reliance on overlapping users or items across domains, or unrealistic assumptions that ignore privacy constraints. In this work, we propose a new framework, \textbf{\emph{MergeRec}}, based on \emph{model merging} under a new and realistic problem setting termed \emph{data-isolated cross-domain sequential recommendation}, where raw user interaction data cannot be shared across domains.
MergeRec consists of three key components: (1) \emph{merging initialization}, (2) \emph{pseudo-user data construction}, and (3) \emph{collaborative merging optimization}. First, we initialize a merged model using training-free merging techniques. Next, we construct pseudo-user data by treating each item as a virtual sequence in each domain, enabling the synthesis of meaningful training samples without relying on real user interactions. Finally, we optimize domain-specific merging weights through a joint objective that combines a \emph{recommendation loss}, which encourages the merged model to identify relevant items, and a \emph{distillation loss}, which transfers collaborative filtering signals from the fine-tuned source models. Extensive experiments demonstrate that MergeRec not only preserves the strengths of the original models but also significantly enhances generalizability to unseen domains. Compared to conventional model merging methods, MergeRec consistently achieves superior performance, with average improvements of up to 17.21\% in Recall@10, highlighting the potential of model merging as a scalable and effective approach for building universal recommender systems. The source code is available at \href{https://github.com/DIALLab-SKKU/MergeRec/}{\texttt{github.com/DIALLab-SKKU/MergeRec}}.

\end{abstract}

\maketitle

\section{Introduction}

\begin{figure*}
\centering
\includegraphics[width=0.96\linewidth]{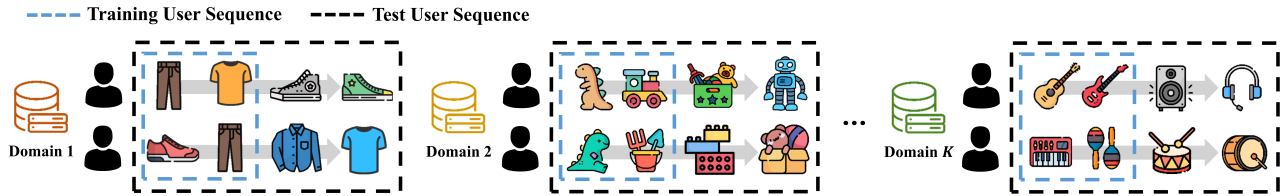}
\Description[Training and test user sequences across domains]%
{Across multiple domains, each user has a training sequence that is fully contained within a longer test sequence. 
The figure shows that test-time user histories extend training interactions with additional future items, indicating that test data are supersets of training data in sequential recommendation.}
\vspace{-1.1mm}
\caption{Illustration of training and test user sequences in sequential recommendation across multiple domains. Each test user sequence (black box) contains all previous interactions, including those from the training period (blue box), highlighting that test data are a superset of training data in real-world scenarios.}

\label{fig:testing_and_pseudo}
\vspace{-1mm}
\end{figure*}

Sequential recommendation (SR) aims to predict the next items a user is likely to prefer based on their interaction history. Recent neural SR models~\cite{KangM18SASRec, SunLWPLOJ19BERT4Rec, HidasiKBT15GRU4Rec, LiZL0WG22MLP4Rec, LiZZW0WG23AutoMLP} employ various architectures to effectively capture sequential dependencies among items. However, they still face inherent challenges such as the cold-start and data sparsity problems~\cite{MoonJCCSL23, MoonKL23, Moon0L25, park2025dan}, which limit their generalizability and overall performance.

Cross-domain sequential recommendation (CDSR) has emerged as a promising research direction~\cite{0002XP0024}. CDSR aims to improve recommendation accuracy by either jointly training models across multiple domains~\cite{ZhuC0LZ19, ParkKCHYCLRYCC23, ParkKYHYCCC24, WangYZZC25} or by transferring knowledge from data-rich domains to sparser ones~\cite{CaoCSLW22, MaRCRZLMR22, LinGZCNSGLJLW24, 00020W0Z0LH025}. However, existing CDSR works face three fundamental limitations. (1) \textbf{User/Item overlap dependency}: knowledge transfer typically relies on the presence of overlapping users or items across domains. However, such overlap is extremely limited in practice. We observe that only 16 users and 0 items are shared across eight Amazon domains, reflecting the real-world nature of independently operated domains; (2) \textbf{Data isolation}: in real-world scenarios, access to raw user data is often restricted due to organizational boundaries or privacy regulations~\cite{YangTZCY20, abs-2102-04925, YangPW00PW0F24}. User logs contain sensitive information and cannot be shared across domains due to privacy restrictions, making domain-specific training data inaccessible; (3) \textbf{Low scalability}: joint training across multiple domains incurs substantial computational overhead, making it impractical for large-scale deployment. Consequently, most prior work has been limited to integration of only two or three domains, leaving scalable multi-domain integration largely unresolved.

We suggest that \emph{model merging}~\cite{IlharcoWGSHKFS22AvgMer1, WortsmanIKLKRLH22AvgMer2, IlharcoRWSHF23TaskArith, YadavTCRB23TIES, YangW00G0T24AdaMerging, JimenezFF23, WangDOFF24Consensus, ZhangAOHA24aTLAS, XuYWWSS24MuDSC, StoicaBBRHH24ZipIt, Yang0WG00T24Surgery, HuangY000O24EMRMerging, 0002LLJGYLGTH024PCBMerging, LuF0QC024TwinMerging, ShirafujiTT25BiasVector, JinHXS025, YoshidaNHYSSMN25, GargiuloCBSSR25TSV} offers an effective alternative for building universal recommender systems. Model merging integrates fine-tuned parameters from multiple domain- or task-specific models into a single unified model. This paradigm provides several advantages that directly address the key limitations of CDSR: (1) It eliminates the need for overlapping users or items across domains; (2) It naturally preserves user privacy, as only model parameters, not sensitive interaction data, are required; (3) It achieves high scalability by avoiding the computational burden of cross-domain joint training.

In this paper, we explore the feasibility of applying model merging to CDSR under a new, realistic problem setting termed \emph{data-isolated CDSR}. This setting is motivated by practical real-world constraints, where user interaction data can be used only to train domain-specific models and cannot be shared across domains or accessed afterward. Unlike conventional CDSR, which often relies on strong and impractical assumptions (\ie, overlapping users or items), data-isolated CDSR allows domains to be disjoint. Moreover, while privacy-preserving CDSR typically requires access to domain-specific interaction data during model optimization, data-isolated CDSR constructs a universal cross-domain recommender system without accessing any user interaction data, thereby providing a stronger guarantee of user privacy.

Under this setting, however, directly applying existing model merging methods is non-trivial for two key reasons. First, since interaction data are not shared across domains, test-time adaptation schemes, commonly used in the model merging paradigm to optimize merging weights, cannot be applied. Second, even if test data were accessible, leveraging test sequences in sequential recommendation would violate the core assumptions of model merging. While the model merging paradigm explicitly prohibits using training data, these assumptions do not hold in sequential recommender systems.
In such systems, test sequences are not independent of the training data but are generated from the same evolving user behavior. Thus, using test-time user interaction sequences during the merging process would inevitably expose training information (Figure~\ref{fig:testing_and_pseudo}). Consequently, leveraging test data for model merging is fundamentally incompatible with the data-isolated CDSR setting.

To this end, we propose \emph{\textbf{MergeRec}}, a novel framework tailored for data-isolated CDSR. MergeRec comprises three key components: (1) \emph{merging initialization}, (2) \emph{pseudo-user data construction}, and (3) \emph{collaborative merging optimization}. First, we synthesize an initial merged model using training-free merging methods based on \emph{task vectors}, defined as the parameter difference between a fine-tuned model and its corresponding pre-trained model, to capture domain-specific knowledge~\cite{IlharcoRWSHF23TaskArith}. Next, we construct pseudo-user data by treating each item in every domain as an individual sequence. Despite its simplicity, MergeRec enables the construction of meaningful samples for merging domains without relying on real user data, effectively simulating cold-start users across domains. Finally, we refine domain-specific merging weights through a recommendation-oriented merging objective.

To design an effective objective function for merging recommender systems, we argue that an ideal merged model should satisfy two fundamental requirements. First, it should be able to decode users' multiple intents, which are often reflected in domain-specific sequential patterns. Second, the unified model should exhibit strong ranking ability, accurately prioritizing items with the highest click probability within each domain context. We point out that existing adaptive merging methods, \ie, \emph{AdaMerging}, address only the latter aspect and are therefore insufficient for merging recommender systems (Section~\ref{sec:proposed_method}).

To overcome this limitation, we propose a joint objective that combines: (1) a \emph{distillation} loss, which leverages the prediction distributions of fine-tuned models as soft labels, and (2) a \emph{recommendation} loss, which treats the top-1 predicted item from each fine-tuned model for a pseudo-user in its corresponding domain as a hard label. The distillation loss transfers collaborative filtering (CF)~\cite{ParkYLPL23MAWU} signals from the fine-tuned models to the merged model, and the recommendation loss guides the merged model to accurately rank items according to their likelihood of being clicked.

Extensive experiments demonstrate that MergeRec not only preserves the strengths of the individual source models but also generalizes effectively to unseen domains. Compared with existing merging methods and strong baselines, including fine-tuned and joint learning models, MergeRec consistently achieves superior performance. Specifically, MergeRec outperforms joint learning and AdaMerging by average gains of 8.72\% and 17.21\% on Recall@10, respectively. These results highlight that model merging can be a scalable and efficient paradigm for building universal recommender systems.

Our contribution can be summarized as follows:
\begin{itemize}[leftmargin=5mm]
    \item \textbf{Thorough empirical analysis}: We provide empirical evidence demonstrating that entropy-based optimization, though effective in computer vision and natural language processing, fundamentally fails to capture the multi-intent behavioral patterns inherent in recommender systems.

\vspace{0.5mm}
    \item \textbf{The first model merging framework for recommender systems}: We propose \emph{\textbf{MergeRec}}, a task vector-based model merging framework tailored for recommender systems. MergeRec comprises three key components: (1) training-free merging initialization, (2) privacy-preserving pseudo-user data construction, and (3) a recommendation-oriented merging objective.

\vspace{0.5mm}
    \item \textbf{Comprehensive evaluation}: Through extensive experiments across eight Amazon benchmark datasets and four backbone architectures, we demonstrate that MergeRec consistently outperforms existing model merging baselines. Notably, MergeRec exhibits superior generalizability to unseen domains and robust performance under data-scarce conditions.
\end{itemize}

\section{Preliminaries}

\subsection{Cross-domain Sequential Recommendation}
Let $\mathcal{D}=\{D_1, D_2, ..., D_K\}$ denote the set of all recommendation domains, where $D_k$ denotes the $k$-th domain. Each domain $D_k$ consists of a set of items $\mathcal{I}_k$ and users $\mathcal{U}_k$. For an arbitrary user $u \in \mathcal{U}_k$, the interaction history is represented as an ordered sequence of items based on timestamps: $u = \left[ i_{1},\, i_{2},\, \ldots,\, i_{|u|} \right]$, where $|u|$ denotes the number of interactions of user $u$.
CDSR models aim to predict and rank items in $\mathcal{I}_k$ by estimating the probability that user $u$ will interact with each item next, conditioned on the user's past interactions:
\begin{equation}
    \mathbf{\theta}^* = \arg\max_\mathbf{\theta} P \bigl(i = i_{|u|+1} \mid u, \mathbf{\theta}\bigr),
\end{equation}
where $\mathbf{\theta}$ denotes the parameters of the CDSR model.

\subsection{Text-based Sequential Recommendation}
Text-based SR~\cite{LiWLFSSM23recformer, hou2024blair} leverages pre-trained language models (PLMs) to encode item-level textual information. By representing both users and items through textual descriptions, this approach enables recommendations for previously unseen (\ie, cold-start) items.

Formally, the textual representation of an item $t_i$ is constructed from its attribute descriptions (\eg, title, brand, and category). The textual representation of a user $t_u$ is defined as the concatenation of the textual representations of all items the user has interacted with:
\begin{equation}
    t_{u} = \left[ t_{i_{1}};\, t_{i_{2}};\, \ldots;\, t_{i_{|u|}} \right],
\end{equation}
where $;$ denotes the concatenation operator.

\begin{figure*}
  \centering
  \includegraphics[width=\linewidth]{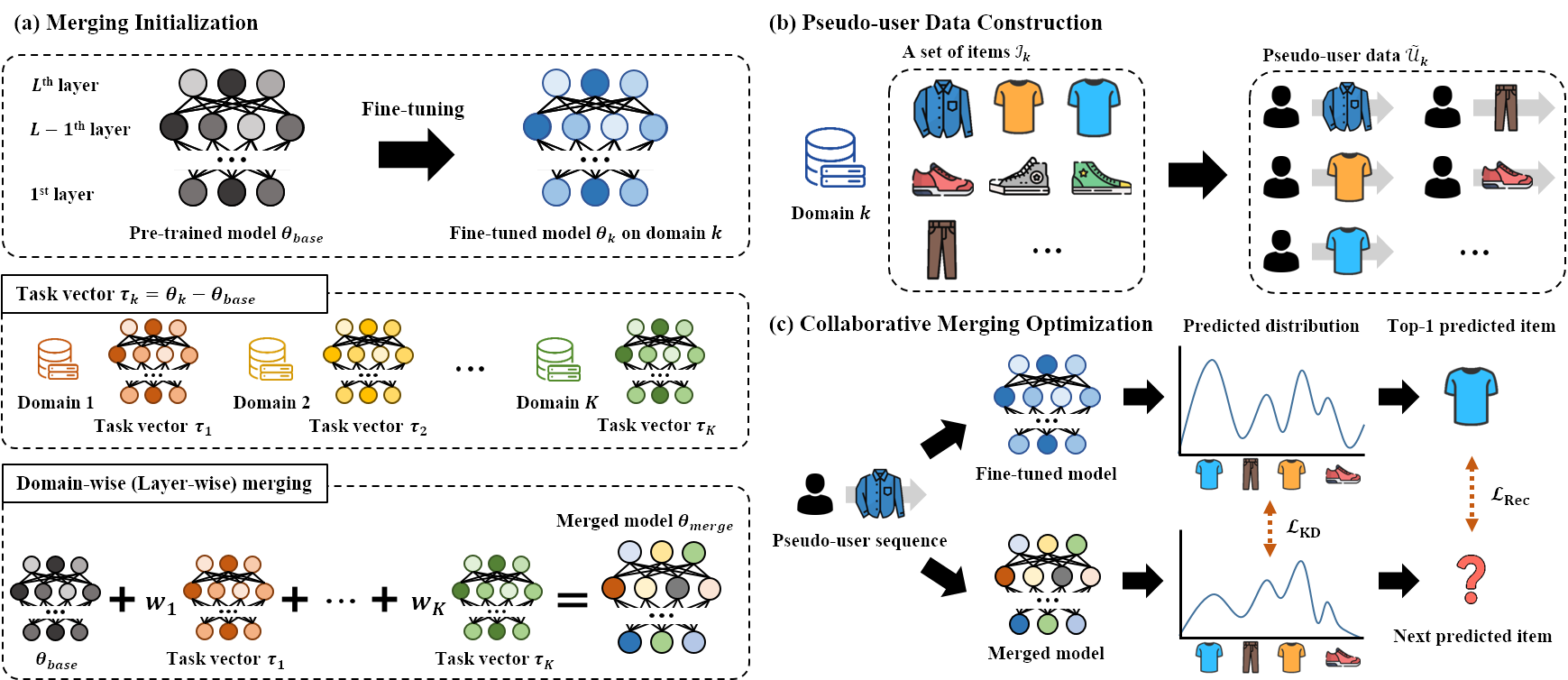}
  \vspace{-6.8mm}
  \caption{Overview of MergeRec with three main components. (a) Merging initialization integrates into a unified model containing multi-domain knowledge. (b) Pseudo-user data construction creates a single-item sequence. (c) Collaborative merging optimization jointly optimizes the recommendation loss $\mathcal{L}_{Rec}$ and the distillation loss $\mathcal{L}_{KD}$. }
  \label{fig:proposed_method}
\vspace{-2mm}
  
\end{figure*}

Let $f(\cdot \mid \mathbf{\theta_k})$ denote a PLM-based encoder with parameters $\mathbf{\theta}_k \in \mathbb{R}^P$ fine-tuned on domain $D_k$, where $P$ is the total number of model parameters. Given the textual inputs $t_u$ and $t_i$, the encoder produces a user representation vector $\mathbf{r}_u \in \mathbb{R}^d$ and an item representation vector $\mathbf{r}_i \in \mathbb{R}^d$ by extracting the final hidden state representations:
\begin{equation}
    \mathbf{r}_u = f\left( t_u \mid \mathbf{\theta}_k \right), ~~ \mathbf{r}_i = f\left( t_i \mid \mathbf{\theta}_k \right),
\end{equation}
where $d$ denotes the dimension of the final hidden representations.

The recommendation score $\hat{y}_{ui}$ between user $u$ and item $i$ is computed as the cosine similarity between their representation vectors:
\begin{equation}
    \hat{y}_{ui} = \cos(\mathbf{r}_u, \mathbf{r}_i).
\end{equation}

The model parameters $\mathbf{\theta}_k$ are optimized using a cross-entropy objective:
\begin{equation}
    \mathbf{\theta}_k^{*} = \arg\min_{\mathbf{\theta}_k} \, \sum_{u \in \mathcal{U}_k} \left( \log{\hat{y}_{k, ui^+}} + \sum_{i^- \in \mathcal{I}_k} \log{(1 - \hat{y}_{k, ui^-})} \right),
\end{equation}
where $i^+$ denotes the next item in the user sequence, and $i^-$ represents negative items in the domain $\mathcal{I}_k$ excluding $i^+$. Note that the domain-specific parameters $\mathbf{\theta}_k$ are initialized from the pre-trained base model parameters $\mathbf{\theta}_{base} \in \mathbb{R}^P$.

\section{Proposed Method: MergeRec} \label{sec:proposed_method}

As illustrated in Figure~\ref{fig:proposed_method}, we design the \emph{\textbf{MergeRec}} framework to address the practical constraint that no interaction data can be shared across domains, termed \emph{data-isolated CDSR}. MergeRec consists of three key components: (1) \emph{Merging Initialization}, which consolidates multiple domain-specific fine-tuned models into a single unified model that integrates knowledge across domains; (2) \emph{Pseudo-user Data Construction}, which synthesizes meaningful merging samples without relying on real user interactions; and (3) \emph{Collaborative Merging Optimization}, which jointly optimizes a recommendation loss and a knowledge distillation loss to enable recommendation-aware parameter integration. Through this design, MergeRec effectively preserves domain-specific CF signals while ensuring strong generalizability across multiple domains.

\begin{table}
\caption{Categorization of cross-domain sequential recommendation problem settings.}\label{tab:research_categorization}
\vspace{-1mm}
\renewcommand{\arraystretch}{1}
\resizebox{\linewidth}{!}{
\begin{tabular}{c|ccc}
\toprule
Setting & \makecell{No User/Item \\ Overlap Required} & Privacy-Aware & Data-Isolated \\
\midrule
\makecell{Conventional \\ CDSR~\cite{CaoCSLW22, 0002XP0024, ParkKYHYCCC24, LinGZCNSGLJLW24, 00020W0Z0LH025}} & \xmark & \xmark & \xmark \\
\makecell{Privacy-preserving \\ CDSR~\cite{abs-2503-23026, TianXCLZ24, WangSZWX24, YangPW00PW0F24, WuLTRWX22}} & \xmark & $\blacktriangle$ & \xmark \\
\midrule
\makecell{Data-isolated \\ CDSR (Proposed)} & \cmark & \cmark & \cmark\\
\bottomrule
\end{tabular}}
\vspace{-1.4mm}
\end{table}

\subsection{Data-isolated CDSR}
We formalize a new and realistic setting for cross-domain sequential recommendation, termed \emph{data-isolated CDSR}. As shown in Table~\ref{tab:research_categorization}, this setting is characterized by two key requirements: (i) it assumes no overlap across domains, (ii) it prohibits access to user interaction data during cross-domain model construction, in contrast to conventional CDSR. Under data-isolated CDSR, domains may be entirely disjoint in both users and items (\eg, $\mathcal{U}_k \cap \mathcal{U}_{k'}=\emptyset$ and $\mathcal{I}_k \cap \mathcal{I}_{k'}=\emptyset$ for $k\neq k'$), reflecting real-world environments that are independently operated. Moreover, this setting enforces a strict \emph{data isolation} constraint: raw interaction logs are accessible only within each domain for training domain-specific models and cannot be shared across domains. Consequently, a cross-domain recommender system must be constructed \emph{without} accessing any domain-specific interaction data. Instead, we assume access only to $K$ domain-specific fine-tuned models $\{\theta_k\}_{k=1}^K$, while the datasets used to train these models remain completely inaccessible. The goal of data-isolated CDSR is to produce a single \emph{universal} sequential recommender system that can operate across multiple domains.

\subsection{Merging Initialization}
\textbf{Problem definition}. Let $f_{\theta_k}(u_k) \rightarrow \hat{y}_k$ denote an SR model fine-tuned on the private data of domain $D_k = \{\mathcal{U}_k, \mathcal{I}_k\}$. For an arbitrary user interaction sequence $u_k \in \mathcal{U}_k$, the model outputs a click probability vector $\hat{\mathbf{y}}_k \in \mathbb{R}^{|\mathcal{I}_k|}$ over candidate items. Without loss of generality, we assume that the model parameters are composed of $L$ layers, \ie, $\theta = \left\{ \theta^1, \theta^2, ..., \theta^L \right\}$.

\vspace{0.5mm}
\noindent
\textbf{Task vector}.
A task vector represents the parameter shift required to adapt a pre-trained model to a specific downstream task~\cite{IlharcoRWSHF23TaskArith, YadavTCRB23TIES, YangW00G0T24AdaMerging, JimenezFF23, WangDOFF24Consensus, ZhangAOHA24aTLAS, HuangY000O24EMRMerging, 0002LLJGYLGTH024PCBMerging, ShirafujiTT25BiasVector, JinHXS025, YoshidaNHYSSMN25}. In our context, each task corresponds to \emph{recommendation within a specific domain}. Accordingly, the task vector captures domain-specific knowledge, enabling a pre-trained model to specialize in that domain.

Formally, the task vector $\boldsymbol{\tau}_k \in \mathbb{R}^P$ for domain $k$ is defined as the difference between the parameters of the fine-tuned model $\mathbf{\theta}_k$ and those of the original pre-trained base model $\mathbf{\theta}_{base}$:
\begin{equation}
\boldsymbol{\tau}_k = \mathbf{\theta}_k - \mathbf{\theta}_{base}.
\end{equation}
where $P$ denotes the total number of model parameters.

\vspace{0.5mm}
\noindent
\textbf{Domain-wise merging}.
Domain-wise merging integrates multiple fine-tuned models by combining their task vectors, each weighted by a domain-specific scalar $w_k$, and adding them to the base model parameters. The merging weights $\mathbf{w}=\{w_1, ..., w_K\}$ can be either uniformly assigned~\cite{IlharcoRWSHF23TaskArith, YadavTCRB23TIES} or adaptively learned to reflect domain characteristics~\cite{YangW00G0T24AdaMerging}. Intuitively, domains containing more distinctive knowledge may receive higher weights, while those sharing similar CF signals may be down-weighted. Formally, domain-wise merging is defined as:
\begin{equation} \label{eq:domain_wise_merging}
    \theta_{merge}=\theta_{base}+\sum_{k=1}^{K}{w_k \cdot \boldsymbol{\tau}_k}.
\end{equation}

In this work, we learn the domain-specific weights $\mathbf{w}$ in a data-driven manner.

\vspace{0.5mm}
\noindent
\textbf{Layer-wise merging}. Since different layers in deep neural networks capture different levels of abstraction~\cite{AkenWLG19, RogersKR20}, applying a single scalar weight per domain may be insufficient to control inter-domain interference. To enable fine-grained integration, we assign independent merging weights to each layer for every domain.

Let $\theta_k = \{\theta_k^1, \dots, \theta_k^L\}$ denote the parameters of the fine-tuned model for domain $k$. The corresponding layer-wise task vector is defined as $\boldsymbol{\tau}_k = \{\theta_k^1 - \theta_{\text{base}}^1, ..., \theta_k^L - \theta_{\text{base}}^L\}$. The layer-wise merging is then defined as:
\begin{equation} \label{eq:layer_wise_merging}
    \theta_{merge}=\left\{ \theta_{base}^l+\sum_{k=1}^{K}{w_k^l \cdot \tau_k^l} \right\}_{l=1}^L.
\end{equation}

The layer-specific merging weights $\mathbf{w}_k = \{w_k^1, \dots, w_k^L\}$ are similarly learned in a data-driven manner.

\subsection{Pseudo-user Data Construction}
User logs in recommender systems typically contain sensitive personal information and cannot be shared across domains, making it challenging to construct data for learning merging weights. To address this, we propose a novel pseudo-user data construction strategy that represents each item in a domain as a single-item interaction sequence. Our design is grounded in the idea that CF knowledge is encapsulated within domain-specific models and can be transferred without relying on the real user data on which they were trained. By leveraging pseudo-users as surrogate inputs, our approach enables learning merging weights without access to domain-specific fine-tuning data while strictly preserving data isolation.

Formally, we construct the pseudo-user set for domain $k$ as:
\begin{equation}
    \mathcal{\tilde{U}}_{k} = \left \{ [i] \mid i \in \mathcal{I}_k \right \}.
\end{equation}

These synthesized samples emulate plausible cold-start users in each domain and thus provide meaningful signals for model merging. Although each pseudo-user sequence contains no explicit sequential context, it serves as a probe to elicit rich CF knowledge encoded in the corresponding domain-specific model $f_{\theta_k}$. We therefore employ each domain-specific model as a teacher, whose conditional distribution $P_{\theta_k}(\cdot \mid [i])$ captures the local co-consumption structure around item $i$, \ie, next-item likelihoods. By distilling the merged model to align with these teacher distributions, we effectively transfer domain-specific CF signals. We observe that even single-item pseudo-user sequences are sufficient for effective model merging (Section~\ref{sec:experimental_results}), providing a practical and privacy-preserving foundation for collaborative merging optimization. While extending pseudo-user sequences to longer contexts may further enrich the transferred signals, we leave this promising direction for future work.

\begin{figure}\label{fig:entropy_in_recommender_systems}
\centering
\includegraphics[width=1\linewidth]{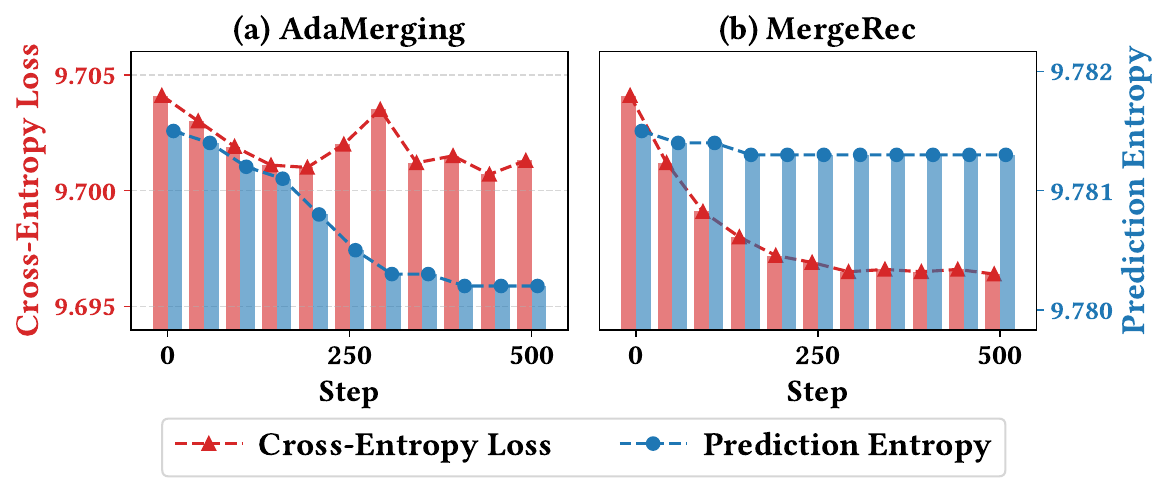}
\vspace{-8mm}
\caption{Cross-entropy loss and prediction entropy dynamics of AdaMerging and MergeRec (Ours) over training steps.}
\label{fig:entropy_loss}
\vspace{-1mm}
\end{figure}

\subsection{Collaborative Merging Optimization}
\textbf{Inadequacy of entropy-based optimization.} We examine a representative adaptive merging method, AdaMerging~\cite{YangW00G0T24AdaMerging}, from the perspective of recommender systems. Figure~\ref{fig:entropy_loss} illustrates the training dynamics of cross-entropy loss and prediction entropy for both AdaMerging and our proposed \emph{\textbf{MergeRec}} over training iterations on eight datasets, evaluated using test user sequences. The cross-entropy loss reflects alignment with the recommendation objective, whereas prediction entropy measures the confidence of the merged model's predictions. As shown in Figure~\ref{fig:entropy_loss} (a), AdaMerging successfully reduces prediction entropy during training but fails to achieve a corresponding decrease in cross-entropy loss. This limitation stems from its exclusive focus on entropy minimization, which merely amplifies confidence in the top-1 predicted item. However, users often exhibit \emph{multi-intent} behavioral patterns rather than a single dominant intent~\cite{TianCNSL22, CenZZZYT20, ZhangYYLFZC022Re4}, making entropy-based optimization alone insufficient to capture the rich CF signals learned by domain-specific fine-tuned models.

To overcome this, we introduce a distillation loss that extends beyond entropy-based optimization by explicitly aligning the merged model with teacher distributions derived from domain-specific models. MergeRec (Figure~\ref{fig:entropy_loss} (b)) simultaneously reduces both cross-entropy loss and prediction entropy, demonstrating more consistent and effective optimization toward the recommendation objective.

\vspace{0.5mm}
\noindent
\textbf{Joint objective function.}
We posit that an ideal merged model for cross-domain recommendation should simultaneously satisfy two essential aspects. First, it should effectively capture diverse user intents reflected in behavioral patterns within each domain and retrieve items relevant to those intents. This requires successfully transferring domain-specific CF knowledge from fine-tuned models to the merged model. Second, the merged model should exhibit strong discriminative capability to accurately identify the items that users are most likely to click on within each domain. To jointly address these requirements, we propose the following optimization function:
\begin{equation}
\mathcal{L} = \mathcal{L}_{\text{Rec}} + \lambda \cdot \mathcal{L}_{\text{KD}},
\end{equation}
where $\lambda$ is a hyperparameter that balances the two losses. In Eqs.~\eqref{eq:domain_wise_merging} and~\eqref{eq:layer_wise_merging}, we optimize only the merging weights $w$, while keeping the base model parameters $\theta_{base}$ and the task vector $\tau$ fixed. Since only $K$ or $K \times L$ domain-specific weights are optimized, MergeRec provides a computationally efficient solution.

The knowledge distillation loss $\mathcal{L}_{\text{KD}}$ integrates domain-specific CF knowledge encoded in fine-tuned models by aligning the predictions of the merged model with those of the corresponding domain-specific models. For a pseudo-user sequence $u \in \mathcal{\tilde{U}}_k$ in domain $k$, we minimize the Kullback-Leibler (KL) divergence between the prediction $\hat{y}_{merge}$ produced by the merged model $\theta_{merge}$ and the prediction $\hat{y}_{k}$ produced by the fine-tuned model $\theta_{k}$:
\begin{equation}
    \begin{split}
    \mathcal{L}_{\text{KD}} &= \sum_{k=1}^{K} \sum_{u \in \mathcal{\tilde{U}}_{k}} \text{KL}(\hat{p}_{merge,u} \parallel \hat{p}_{k,u}) \\
    &= \sum_{k=1}^{K} \sum_{u \in \mathcal{\tilde{U}}_{k}} \sum_{i \in \mathcal{I}_k} \hat{p}_{merge,ui} \log{\frac{\hat{p}_{merge, ui}}{\hat{p}_{k,ui}}},
    \end{split}
\end{equation}
where $\hat{p}_{*}=\text{softmax}(\hat{y}_{*} / T)$, and $T$ denotes the temperature hyperparameter which is empirically set to $1$ in our experiments.

The recommendation loss $\mathcal{L}_{\text{Rec}}$ encourages the merged model to accurately identify items aligned with user intent by assigning high scores to the next item and low scores to others. Since real user sequences are unavailable, we leverage the top-1 predicted item $\tilde{i}^+$, obtained by feeding pseudo-user data into the corresponding fine-tuned model, as a positive pseudo-label:
\begin{equation}
    \begin{split}
    \mathcal{L}_{\text{Rec}} = \sum_{k=1}^{K} \sum_{u \in \mathcal{\tilde{U}}_{k}} \left( \log{\hat{y}_{merge,u\tilde{i}^+}} + \sum_{i^- \in \mathcal{I}_k} \log{(1 - \hat{y}_{merge,ui^-})} \right).
    \end{split}
\end{equation}

\section{Experimental Setup}

\begin{table*}[t]
\caption{Performance comparison with six baseline methods on four backbone models, \ie, RecFormer-base/large~\cite{LiWLFSSM23recformer} and BLaIR-base/large~\cite{hou2024blair}. We report normalized Recall@10 performance (\%) relative to the fine-tuned model's performance, which is 100\%. The best results are marked in \textbf{bold}, and the second-best results are shown as \underline{underlined}. `*’ indicates the statistically significant gain of MergeRec over the best baseline model (p < 0.02 for one-tailed t-test).}
\vspace{-2.7mm}
\centering
\begin{adjustbox}{width=0.95\textwidth}
\label{tab:overall}

\begin{tabular}{c|l|>{\columncolor{gray!10}}rrrrrrrrr}
\toprule
\multicolumn{1}{c|}{\textbf{Backbone}} & 
\multicolumn{1}{c|}{\textbf{Method}} & 
\multicolumn{1}{c}{\textbf{Avg.}} & 
\multicolumn{1}{c}{Arts} & 
\multicolumn{1}{c}{Beauty} & 
\multicolumn{1}{c}{Inst.} & 
\multicolumn{1}{c}{Office} & 
\multicolumn{1}{c}{Pantry} & 
\multicolumn{1}{c}{Sci.} & 
\multicolumn{1}{c}{Sports} & 
\multicolumn{1}{c}{Toys} \\
\midrule

\multirow{9}{*}{RecFormer-base}
& Zero-shot & 75.46 & 76.04 & 63.66 & 72.15 & 61.61 & 74.23 & 92.85 & 71.11 & 84.94 \\
& Joint Learning & 80.17 & 77.37 & 79.53 & 83.26 & 73.37 & 85.34 & 79.96 & 69.62 & 92.94 \\
\cmidrule{2-11}
& Weight Averaging & 89.84 & 91.20 & 83.05 & 90.86 & 77.39 & 93.62 & 99.68 & \textbf{86.66} & 93.00 \\
& Task Arithmetic & 88.95 & 91.52 & 81.45 & 88.12 & 83.79 & 92.67 & 98.90 & 81.32 & 82.34 \\
& TIES & 91.08 & 93.29 & \textbf{85.60} & 90.57 & \textbf{88.22} & 92.75 & 97.68 & \uline{85.56} & 86.25 \\
& AdaMerging (Domain-wise) & 78.91 & 87.26 & 58.66 & 75.81 & 64.78 & 75.02 & 92.56 & 67.97 & \uline{93.11} \\
& AdaMerging (Layer-wise) & 67.36 & 67.17 & 32.92 & 67.36 & 58.54 & 65.64 & 84.54 & 65.24 & 84.93 \\
\cmidrule{2-11}
& MergeRec (Domain-wise) & \textbf{92.33}* & \textbf{96.14}* & 83.69 & \uline{90.93} & \uline{84.27} & \textbf{95.56}* & \uline{100.73}* & 84.00 & \textbf{93.35}* \\
& MergeRec (Layer-wise) & \uline{92.08}* & \uline{95.45}* & \uline{84.85} & \textbf{91.13}* & 83.01 & \uline{94.86}* & \textbf{101.44}* & 85.07 & 92.14 \\
\midrule

\multirow{9}{*}{RecFormer-large}
& Zero-shot & 59.07 & 72.57 & 51.89 & 46.03 & 39.23 & 44.34 & 82.00 & 62.54 & 63.57 \\
& Joint Learning & 83.73 & 83.14 & 79.48 & 79.76 & 73.61 & 90.27 & 92.99 & 91.49 & 83.61 \\
\cmidrule{2-11}
& Weight Averaging & 91.23 & 92.27 & 87.10 & 88.94 & 79.13 & \textbf{96.46} & \textbf{98.32} & \textbf{98.88} & 94.36 \\
& Task Arithmetic & 87.99 & 91.32 & 83.32 & 88.26 & 82.78 & 92.08 & 92.99 & 87.80 & 81.14 \\
& TIES & 89.96 & 93.00 & 85.74 & \uline{90.25} & \textbf{88.17} & 92.91 & 94.10 & 89.09 & 80.79 \\
& AdaMerging (Domain-wise) & 72.59 & 83.37 & 59.49 & 77.55 & 63.74 & 61.37 & 83.96 & 58.14 & 70.69 \\
& AdaMerging (Layer-wise) & 70.80 & 79.59 & 52.28 & 68.27 & 60.64 & 67.12 & 83.08 & 72.97 & 72.01 \\
\cmidrule{2-11}
& MergeRec (Domain-wise) & \textbf{92.99}* & \textbf{95.19}* & \textbf{90.08}* & \textbf{91.75}* & \uline{83.39} & 96.20 & 97.41 & 94.74 & \textbf{96.64}* \\
& MergeRec (Layer-wise) & \uline{92.50}* & \uline{94.21}* & \uline{89.77}* & \uline{90.25} & 82.18 & \uline{96.27} & \uline{97.86} & \uline{96.53} & \uline{96.17}* \\
\midrule

\multirow{9}{*}{BLaIR-base}
& Zero-shot & 41.10 & 45.88 & 35.57 & 31.55 & 27.74 & 47.20 & 54.94 & 34.50 & 42.32 \\
& Joint Learning & 83.67 & 82.89 & \textbf{97.30} & \textbf{83.52} & 73.09 & \uline{91.24} & 74.85 & \textbf{92.88} & \textbf{94.87} \\
\cmidrule{2-11}
& Weight Averaging & \uline{87.90} & 91.98 & \uline{93.97} & 78.41 & 76.12 & \textbf{91.39} & \textbf{99.95} & \uline{89.45} & 81.81 \\
& Task Arithmetic & 61.50 & 57.05 & 53.43 & 66.66 & 61.61 & 62.42 & 73.27 & 49.34 & 53.78 \\
& TIES & 82.95 & 84.37 & 78.02 & 78.26 & \textbf{87.64} & 77.18 & 88.52 & 87.26 & 77.39 \\
& AdaMerging (Domain-wise) & 60.60 & 67.93 & 52.07 & 53.40 & 44.83 & 50.80 & 72.94 & 72.91 & 73.14 \\
& AdaMerging (Layer-wise) & 68.48 & 73.09 & 61.14 & 68.86 & 55.34 & 57.54 & 83.92 & 74.23 & 70.61 \\
\cmidrule{2-11}
& MergeRec (Domain-wise) & 87.40 & \textbf{94.42}* & 89.38 & 77.75 & \uline{81.44} & 85.18 & 95.48 & 85.05 & \uline{84.01} \\
& MergeRec (Layer-wise) & \textbf{88.01} & \uline{93.53}* & 91.83 & \uline{78.98} & 81.15 & 87.82 & \uline{96.55} & 87.45 & 83.06 \\
\midrule

\multirow{9}{*}{BLaIR-large}
& Zero-shot & 33.44 & 38.42 & 29.99 & 24.55 & 19.23 & 46.48 & 37.83 & 27.84 & 41.70 \\
& Joint Learning & 84.59 & 83.33 & 81.01 & 86.78 & 78.56 & 90.60 & 78.48 & 83.85 & \textbf{100.63} \\
\cmidrule{2-11}
& Weight Averaging & 88.99 & 92.76 & 89.13 & 83.82 & 76.25 & \uline{93.20} & \uline{103.56} & 86.36 & 81.69 \\
& Task Arithmetic & 78.86 & 82.64 & 79.11 & 65.86 & 74.43 & 83.64 & 93.56 & 73.13 & 67.87 \\
& TIES & 90.90 & 93.78 & 91.09 & 87.76 & \textbf{90.01} & 92.58 & 98.36 & \textbf{93.67} & 75.35 \\
& AdaMerging (Domain-wise) & 69.27 & 82.21 & 62.36 & 71.76 & 51.16 & 60.90 & 78.41 & 77.27 & 68.45 \\
& AdaMerging (Layer-wise) & 78.22 & 84.78 & 73.80 & 78.57 & 62.43 & 83.01 & 84.04 & 71.51 & 83.00 \\
\cmidrule{2-11}
& MergeRec (Domain-wise) & \uline{91.80}* & \uline{97.80}* & \uline{93.07}* & \textbf{91.81}* & 80.51 & 89.76 & 102.88 & 90.16 & 82.84 \\
& MergeRec (Layer-wise) & \textbf{93.70}* & \textbf{98.73}* & \textbf{95.40}* & \uline{91.74}* & \uline{82.62} & \textbf{94.17}* & \textbf{105.18}* & \uline{90.28} & \uline{85.63} \\

\bottomrule
\end{tabular}
\end{adjustbox}
\vspace{-1mm}
\end{table*}

\noindent
\textbf{Datasets.} To simulate a cross-domain recommendation environment, we use eight categories from the Amazon dataset\footnote{\url{https://cseweb.ucsd.edu/~jmcauley/datasets/amazon/links.html}}~\footnote{\url{https://cseweb.ucsd.edu/~jmcauley/datasets/amazon_v2/}}: Arts, Beauty, Instruments, Office, Pantry, Scientific, Sports, and Toys. Following existing work~\cite{KangM18SASRec, SunLWPLOJ19BERT4Rec}, we adopt a 5-core setting, \ie, users and items with fewer than five interactions are removed. Detailed dataset statistics are provided in Appendix~\ref{app:statistics}.

\noindent
\textbf{Baselines.} We compare MergeRec with the following methods:
\begin{itemize}[leftmargin=5mm]
  \item \textbf{Zero-shot}: Directly applies pre-trained text-based SR models without fine-tuning on a specific domain.
  \item \textbf{Fine-tuning}: Fine-tunes pre-trained models using domain-specific interaction data.
  \item \textbf{Joint Learning}: Trains a unified model on aggregated multi-domain datasets with shared parameters.
  \item \textbf{Task Arithmetic}~\cite{IlharcoRWSHF23TaskArith}: Constructs a cross-domain model by linearly adding task vectors to a pre-trained model.
  \item \textbf{TIES}~\cite{YadavTCRB23TIES}: Reduces noise and conflicts between task vectors by selecting parameters with large variance and aligning their signs.
  \item \textbf{AdaMerging}~\cite{YangW00G0T24AdaMerging}: Learns adaptive merging weights in an unsupervised manner by minimizing the prediction entropy of the merged model.
\end{itemize}

We evaluate all methods on RecFormer-base/large~\cite{LiWLFSSM23recformer}, a representative text-based SR model, and BLaIR-base/large~\cite{hou2024blair}, a language model post-trained on a recommendation corpus, as backbone architectures. For a fair comparison under the data-isolation setting, AdaMerging is adapted to use the same pseudo-user data as MergeRec, treating it as unlabeled inputs for entropy-based optimization. Implementation details are provided in Appendix~\ref{app:implementation}.

\noindent
\textbf{Evaluation protocol.} Following~\cite{KangM18SASRec, SunLWPLOJ19BERT4Rec}, we adopt the leave-one-out strategy to split the train, validation, and test datasets. For each user, the most recently interacted item is used for testing, the second most recently interacted item for validation, and the rest for training. Note that the training data is used only for fine-tuning and joint learning. We evaluate recommendation performance using Recall@10 (R@10) and NDCG@10 (N@10). Following~\cite{YadavTCRB23TIES}, we normalize the performance of each method by that of its corresponding fine-tuned model.
The normalized results are reported in Table~\ref{tab:overall} and Figures~\ref{fig:performance_over_the_number_of_merged_models},~\ref{fig:popularity}, and~\ref{fig:hyperparameter_sensi}.

\section{Experimental Results} \label{sec:experimental_results}

\subsection{Overall Performance} \label{sec:overall_performance}
Table~\ref{tab:overall} shows the normalized R@10 of MergeRec and seven baseline methods across eight datasets and four backbone models, where the performance of the fine-tuned model on each dataset is normalized to 100\%. The corresponding normalized N@10, unnormalized R@10, N@10 results are provided in Appendix~\ref{app:results}.

MergeRec consistently achieves the best average performance across all datasets and backbone models for both the domain-wise and layer-wise variants. Specifically, MergeRec outperforms Joint Learning and AdaMerging with average gains of 8.72\% and 17.21\%, respectively. This indicates that MergeRec simultaneously enhances the ranking discriminative ability of the merged model and effectively transfers domain-specific CF knowledge from fine-tuned models. Meanwhile, AdaMerging performs substantially worse across all datasets and backbone models, suggesting that merely amplifying prediction confidence is insufficient to capture the diverse CF signals present across multiple domains. Furthermore, MergeRec surpasses training-free model merging methods (\ie, Task Arithmetic and TIES) by average gains of 9.90\% and 3.04\%, respectively.

Several model merging methods, \ie, MergeRec, Weight Averaging, and TIES, consistently outperform Joint Learning. These results demonstrate that model merging can effectively capture complementary domain knowledge and improve recommendation quality without relying on cross-domain training data. It further highlights the practical advantages of model merging, as it not only reduces computational overhead but also enables synergistic knowledge transfer across domains without end-to-end re-training.

Cross-domain merging is particularly beneficial for data-scarce domains. On the Scientific dataset, MergeRec achieves improvements of 1.44\% for RecFormer-base (Domain-wise) and 5.18\% for BLaIR-large (Layer-wise) compared to their respective fine-tuned counterparts. These improvements can be attributed to the limited number of users in the Scientific domain, where the merged model benefits more substantially from cross-domain knowledge transferred from other domains.

Overall, these results demonstrate that MergeRec provides a robust and scalable solution for cross-domain model merging in recommender systems, delivering consistent and significant performance gains across diverse domains and backbone architectures.

\begin{table}[t]
\caption{Performance comparison over varying the data sparsity of a target-domain training set. We merge five source models (trained on Arts, Beauty, Pantry, Sports, and Toys) with one target model for each of the three datasets (Inst., Office, and Sci.). Each target model is trained on a subset of the full dataset (1\%, 5\%, and 10\%). `Ratio' denotes the fraction of target-domain training data used for fine-tuning, and the RecFormer-base is used as the backbone. The metric is Recall@10. } \label{tab:scarse_training_data}
\centering
\vspace{-3mm}
\begin{adjustbox}{width=0.95\linewidth}
    \begin{tabular}{c|l|>{\columncolor{gray!10}}rrrr}
    \toprule

    \multicolumn{1}{c|}{\textbf{Ratio}} & 
    \multicolumn{1}{c|}{\textbf{Method}} & 
    \multicolumn{1}{c}{\textbf{Avg.}} & 
    \multicolumn{1}{c}{Inst.} & 
    \multicolumn{1}{c}{Office} & 
    \multicolumn{1}{c}{Sci.} \\
    \midrule
    \multirow{5}{*}{1\%} &
    Fine-tuning       & 0.0989 & 0.0745 & 0.0953 & 0.1268 \\
    & Task Arithmetic     & \uline{0.1069} & \uline{0.0828} & \uline{0.0995} & \uline{0.1385} \\
    & TIES            & 0.0650 & 0.0158 & 0.0614 & 0.1178 \\
    & AdaMerging      & 0.0982 & 0.0732 & 0.0898 & 0.1318 \\
    & MergeRec        & \textbf{0.1089} & \textbf{0.0859} & \textbf{0.1013} & \textbf{0.1394} \\
    \midrule
    \multirow{5}{*}{5\%} &
    Fine-tuning       & 0.1057 & 0.0752 & \uline{0.1094} & 0.1325 \\
    & Task Arithmetic     & \uline{0.1101} & \uline{0.0859} & 0.1050 & \uline{0.1392} \\
    & TIES            & 0.0800 & 0.0812 & 0.1017 & 0.0570 \\
    & AdaMerging      & 0.1047 & 0.0814 & 0.1023 & 0.1304 \\
    & MergeRec        & \textbf{0.1128} & \textbf{0.0888} & \textbf{0.1098} & \textbf{0.1399} \\
    \midrule
    \multirow{5}{*}{10\%} &
    Fine-tuning       & \uline{0.1118} & 0.0838 & \textbf{0.1175} & 0.1341 \\
    & Task Arithmetic     & 0.1107 & \uline{0.0860} & 0.1055 & \textbf{0.1405} \\
    & TIES            & 0.0840 & 0.0809 & 0.1017 & 0.0693 \\
    & AdaMerging      & 0.0982 & 0.0755 & 0.0935 & 0.1258 \\
    & MergeRec        & \textbf{0.1120} & \textbf{0.0887} & \uline{0.1088} & \uline{0.1386} \\

\bottomrule
\end{tabular}
\end{adjustbox}
\vspace{-5mm}
\end{table}

\subsection{Model Merging on Scarce Training Data} \label{sec:sparsity}

Collecting sufficient data is often challenging in the early stages of recommender systems. Under such data-scarce conditions, model merging can offer a promising solution for improving model generalization by leveraging knowledge from data-rich domains. We investigate whether model merging can improve recommendation performance in domains with limited data.

To simulate this scenario, we divide the eight domains into two groups: five source domains (Arts, Beauty, Pantry, Sports, Toys) and three target domains (Instruments, Office, Scientific). We then vary the degree of scarcity in the target domain by randomly sampling $k\%$ of users ($k = 1, 5, 10$) from the whole user set to construct data-scarce training sets. Each fine-tuned model trained on a data-scarce target domain is subsequently merged with the fine-tuned models trained from the five source domains.

Table~\ref{tab:scarse_training_data} compares MergeRec with five merging methods under different levels of data scarcity using the RecFormer-base backbone. MergeRec consistently outperforms the corresponding fine-tuned models, demonstrating its strong ability to transfer knowledge across domains even under extreme data scarcity. Moreover, MergeRec consistently surpasses AdaMerging across all scarcity levels, indicating that our recommendation-oriented optimization captures transferable CF patterns more effectively than the entropy minimization approach. These results confirm that MergeRec can reliably transfer CF signals from data-rich to data-scarce domains.

\begin{table}[t]
\caption{Unseen-domain performance comparison of task-vector based model merging methods using the RecFormer-base backbone. We train a merged model on the Arts, Beauty, Pantry, Sports, and Toys datasets and test it on the Inst., Office, and Sci. datasets. The metric is Recall@10. } \label{tab:unseen_domains}
\vspace{-3.5mm}
\begin{adjustbox}{width=0.85\linewidth}

\centering
\begin{tabular}{l|>{\columncolor{gray!10}}rrrr}
\toprule
\multicolumn{1}{c|}{\textbf{Method}} & 
\multicolumn{1}{c}{\textbf{Avg.}} & 
\multicolumn{1}{c}{Inst.} & 
\multicolumn{1}{c}{Office} & 
\multicolumn{1}{c}{Sci.} \\

\midrule
Task Arithmetic  & \uline{0.1062} & \uline{0.0817} & 0.0984 & 0.1385 \\
TIES             & \uline{0.1062} & \uline{0.0817} & \uline{0.0988} & 0.1380 \\
AdaMerging       & 0.0998 & 0.0759 & 0.0901 & 0.1333 \\
MergeRec         & \textbf{0.1081} & \textbf{0.0849} & \textbf{0.1002} & \textbf{0.1393} \\
\bottomrule
\end{tabular}
\end{adjustbox}
\vspace{-3.4mm}
\end{table}

\subsection{Performance on Unseen Domains} \label{sec:unseen_domain_performance}

To examine whether model merging remains effective under extreme conditions where no interaction data are available for the target domain, we merge models trained on five source domains (Arts, Beauty, Pantry, Sports, Toys) and evaluate their performance on three unseen target domains (Inst., Office, Sci.). Note that neither interaction data nor domain-specific models from the target domains are used in constructing the merged model.

Table~\ref{tab:unseen_domains} shows the performance of four merging methods on unseen domains. MergeRec consistently outperforms the other methods, achieving improvements of up to 3.92\% over Task Arithmetic. These results demonstrate the strong generalizability of MergeRec and highlight its effectiveness in transferring knowledge to entirely unseen domains without relying on any target-domain data.

\begin{figure}
\centering
\includegraphics[width=0.9\linewidth]{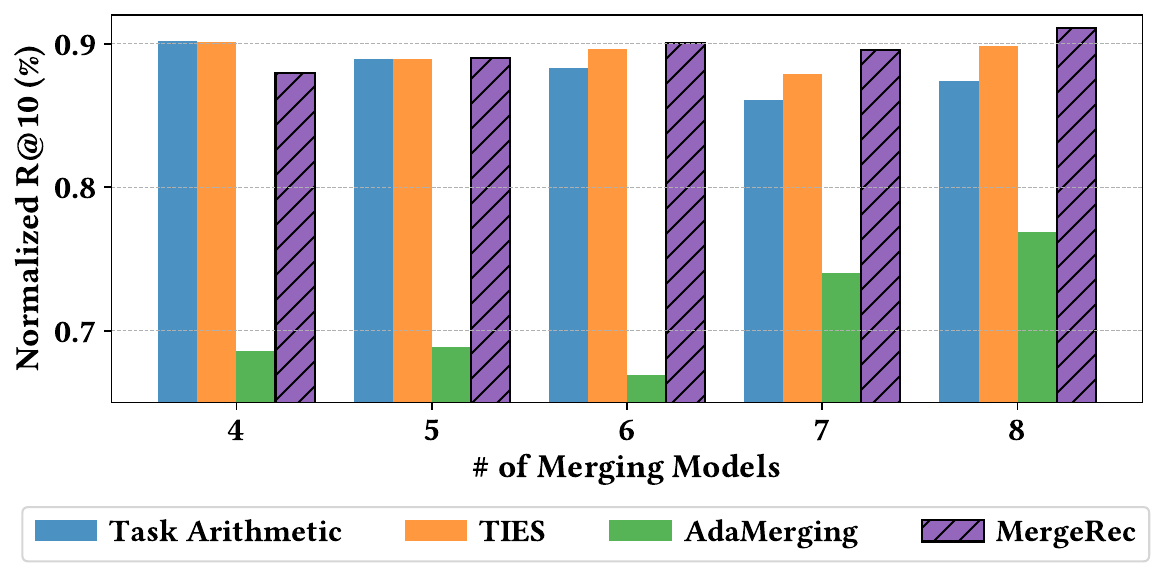}
\vspace{-4mm}
\caption{Normalized average performance across varying the number of datasets for model merging.}
\label{fig:performance_over_the_number_of_merged_models}
\vspace{-1mm}
\end{figure}
\begin{figure}
\centering
\includegraphics[width=1\linewidth]{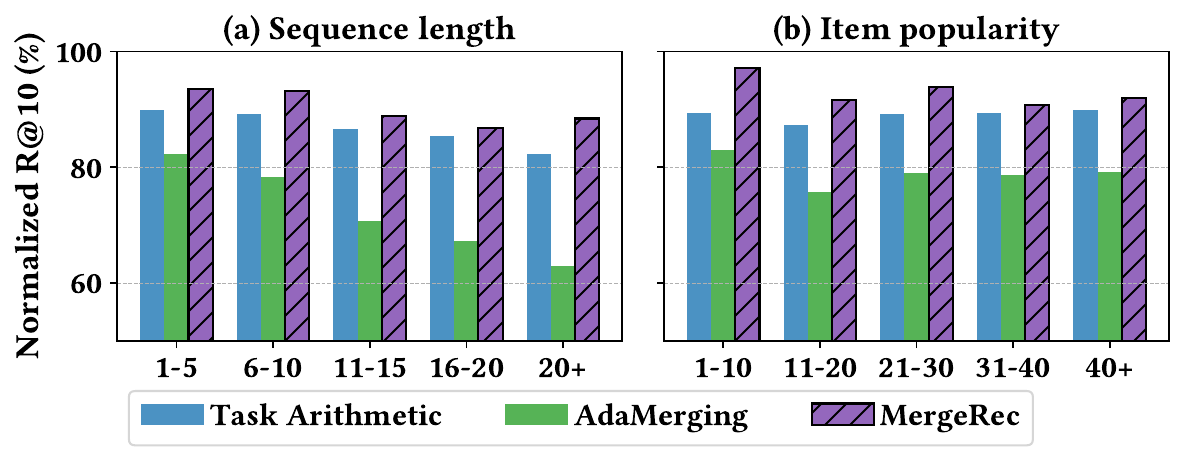}
\vspace{-8mm}
\caption{Normalized average performance across user and item groups on eight datasets. User and item groups are divided by sequence length and item popularity, respectively.}
\label{fig:popularity}
\vspace{-1mm}
\end{figure}
\begin{figure}
\centering
\includegraphics[width=0.9\linewidth]{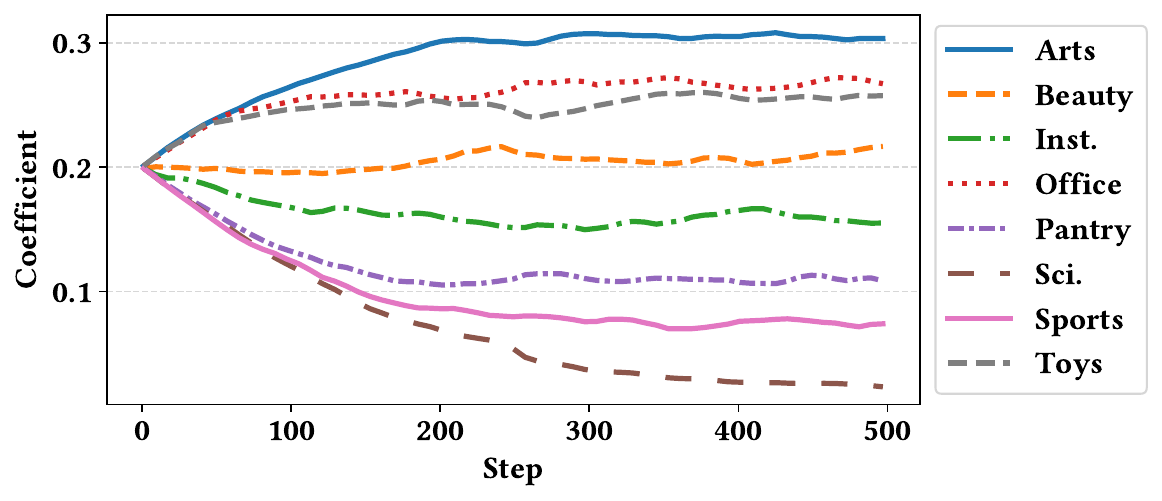}
\vspace{-4mm}
\caption{Domain-specific coefficients dynamics over training steps on the RecFormer-base backbone.}
\label{fig:model_coefficients_over_training}
\vspace{-3mm}
\end{figure}

\subsection{Further Analysis} \label{sec:further_analysis}
\textbf{Effect of the number of merged domains}.
To assess the robustness of MergeRec, we analyze how recommendation performance varies with the number of merged models, as shown in Figure~\ref{fig:performance_over_the_number_of_merged_models}.
Task Arithmetic and TIES exhibit competitive performance when merging a small number of domains (four or five), but their performance deteriorates as more domains are included, eventually falling behind MergeRec. AdaMerging consistently performs poorly across all settings, indicating that its entropy-based optimization strategy fails to effectively adapt to recommendation tasks.
In contrast, MergeRec maintains stable performance regardless of the number of merged domains. Notably, its average performance improves as more domains are integrated, highlighting its strong ability to effectively consolidate and leverage knowledge across an increasing number of diverse domains.

\vspace{0.5mm}
\noindent
\textbf{Performance across user and item groups}.
To further analyze the sources of performance improvements, we partition the test data into multiple groups based on user history length and target item popularity. Figure~\ref{fig:popularity} shows the performance of three merging methods across five sequence-length groups and five item popularity ranges. All methods are evaluated on the eight datasets, and the results are averaged and normalized by the performance of the corresponding fine-tuned model within each group.

As shown in Figure~\ref{fig:popularity} (a), AdaMerging exhibits severe performance degradation as the sequence length increases. This result indicates that AdaMerging struggles to capture CF signals in longer sequences, as it does not learn inter-item relationships during optimization. In contrast, MergeRec consistently achieves the best performance across all sequence lengths by effectively transferring inter-item relationships through the proposed objective function. Figure~\ref{fig:popularity} (b) shows that MergeRec outperforms other methods across all item popularity ranges. Notably, in the least popular item group (1–10), MergeRec achieves substantial performance gains over competing methods, highlighting its ability to leverage cross-domain knowledge to recommend less popular items.

\vspace{0.5mm}
\noindent
\textbf{Domain-specific weight dynamics}.
To understand how the model adapts to different domains during merging, we analyze the training trajectories of the domain-wise merging weights $\mathbf{w}$ across eight domains using the RecFormer-base model. Figure~\ref{fig:model_coefficients_over_training} illustrates the evolution of each domain's weight during training. We observe that the weights gradually converge to distinct values for each domain, indicating that the model learns to differentiate the relative importance of individual domains. Notably, domains with larger scales (\eg, Arts and Office) converge to higher weight coefficients. These results suggest that domains exhibiting more complex CF patterns are assigned greater emphasis, enabling the merged model to better preserve domain-specific knowledge. The corresponding results for other backbone models are reported in Appendix~\ref{domain_weights_other_backbones}.

\begin{figure}
\centering
\includegraphics[width=0.95\linewidth]{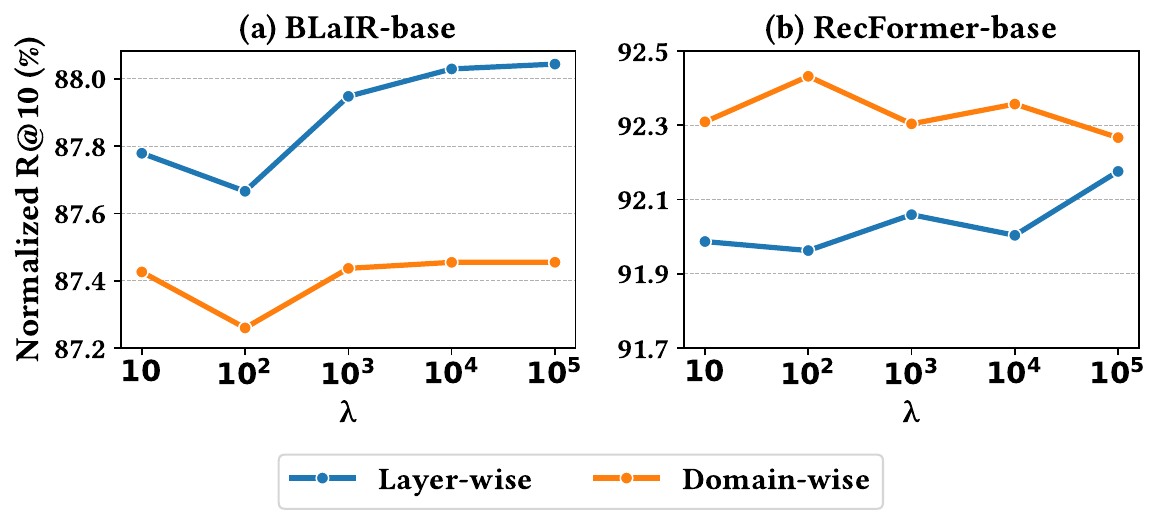}
\vspace{-4mm}
\caption{Normalized average performance over varying $\lambda$.} \label{fig:hyperparameter_sensi}
\vspace{-3mm}
\end{figure}

\subsection{Hyperparameter Sensitivity} \label{sec:hyperparam_sensi}
Figure~\ref{fig:hyperparameter_sensi} depicts the effect of the trade-off hyperparameter $\lambda$ on recommendation performance. The results are averaged across all eight datasets and normalized by the performance of the corresponding fine-tuned models. We find that performance generally improves as $\lambda$ increases, except for domain-wise merging with the RecFormer-base backbone. This suggests that placing greater emphasis on the knowledge distillation loss is typically more effective.
This can be attributed to the model learning the full item-prediction distribution from domain-specific teacher models, which enables the merged model to capture richer and more informative CF signals.
In contrast, methods that focus solely on amplifying top-1 prediction confidence provide a more limited supervisory signal.
\section{Related Work}

\subsection{Cross-Domain Sequential Recommendation}
CDSR~\cite{0002XP0024} has emerged as an effective approach to alleviate data sparsity and cold-start problems in single-domain recommender systems~\cite{KangM18SASRec, SunLWPLOJ19BERT4Rec, HidasiKBT15GRU4Rec, LiZL0WG22MLP4Rec, LiZZW0WG23AutoMLP}. Existing CDSR approaches can be categorized into three types: single-target CDSR, dual-target CDSR, and multi-target CDSR. Additionally, privacy-preserving cross-domain recommendation (CDR) has gained increasing attention in privacy-sensitive and federated learning scenarios.

\vspace{0.5mm}
\noindent
\textbf{Single-target CDSR}.
Single-target CDSR~\cite{CaoCSLW22, MaRCRZLMR22, LinGZCNSGLJLW24, 00020W0Z0LH025} is the most extensively studied setting, aiming to improve recommendation performance in a data-scarce target domain by transferring knowledge from data-rich source domains with abundant user-item interactions or auxiliary information.
C$^2$DSR~\cite{CaoCSLW22} jointly models intra-sequence and inter-sequence item relationships through graph neural networks and self-attention mechanisms.
To enhance self-attention modules, MAN~\cite{LinGZCNSGLJLW24} introduces both local and global attention modules to capture domain-specific and cross-domain information.
Recently, LLM4CDSR~\cite{00020W0Z0LH025} leverages large language models to generate semantic item representations from textual attributes and hierarchical user profiles from interaction sequences, facilitating cross-domain knowledge transfer.

\vspace{0.5mm}
\noindent
\textbf{Dual- and multi-target CDSR}.
Dual- and multi-target CDSR~\cite{ZhuC0LZ19, ParkKCHYCLRYCC23, ParkKYHYCCC24, WangYZZC25} aims to simultaneously improve recommendation performance across multiple domains. DTCDR~\cite{ZhuC0LZ19} adopts a multi-task learning framework that bidirectionally transfers user preference representations across domains via shared embeddings. Recent studies~\cite{ParkKCHYCLRYCC23, ParkKYHYCCC24} have highlighted the negative transfer problem, where knowledge from other domains can fail to provide beneficial contributions. To address this, CGRec~\cite{ParkKCHYCLRYCC23} and SyNCRec~\cite{ParkKYHYCCC24} introduce adaptive loss weighting based on estimated transfer gaps between single-domain and cross-domain sequential recommendation tasks. However, these approaches fundamentally rely on overlapping users or items to enable knowledge transfer and do not account for the privacy constraints commonly required in real-world scenarios.

\vspace{0.5mm}
\noindent
\textbf{Privacy-preserving CDR}.
Privacy-preserving CDR~\cite{abs-2503-23026, TianXCLZ24, WangSZWX24, YangPW00PW0F24, WuLTRWX22} aims to improve recommendation performance under settings where access to raw data from individual domains is restricted or entirely unavailable. An early study~\cite{WuLTRWX22} separates personalized and transferable components to enable privacy-compliant recommendations.
P2M2-CDR~\cite{WangSZWX24} disentangles domain-common and domain-specific embeddings while applying local differential privacy to perturb shared representations.
FedGCDR~\cite{YangPW00PW0F24} further adopts federated graph learning with differential privacy-based knowledge extraction and graph expansion to mitigate the negative transfer problem.
However, these approaches typically rely on overlapping users across domains and require cross-domain coordination during training, where domain-specific data are used alongside data from other domains (\eg, exchanged model parameters or gradients). Therefore, they do not fully satisfy the \emph{data isolation} constraint.

In this paper, \emph{we pioneer the application of model merging to data-isolated multi-target CDSR}, providing a scalable solution for integrated multi-domain recommendations without requiring direct access to domain-specific user interaction data, thereby preserving user privacy.

\subsection{Model Merging}
Model merging~\cite{MatenaR22Fisher, IlharcoWGSHKFS22AvgMer1, WortsmanIKLKRLH22AvgMer2, IlharcoRWSHF23TaskArith, YadavTCRB23TIES, YangW00G0T24AdaMerging, JimenezFF23, WangDOFF24Consensus, ZhangAOHA24aTLAS, XuYWWSS24MuDSC, StoicaBBRHH24ZipIt, Yang0WG00T24Surgery, HuangY000O24EMRMerging, 0002LLJGYLGTH024PCBMerging, LuF0QC024TwinMerging, ShirafujiTT25BiasVector, JinHXS025, YoshidaNHYSSMN25, GargiuloCBSSR25TSV} aims to improve the generalization of domain-specific fine-tuned models by consolidating knowledge from multiple models trained on diverse domains or tasks into a single model, typically without requiring access to training data. The simplest approaches~\cite{IlharcoWGSHKFS22AvgMer1, WortsmanIKLKRLH22AvgMer2} perform parameter averaging across fine-tuned models that share the same backbone pre-trained model.

Beyond these early methods, \emph{task vector}-based model merging~\cite{IlharcoRWSHF23TaskArith, YadavTCRB23TIES, YangW00G0T24AdaMerging, JimenezFF23, WangDOFF24Consensus, ZhangAOHA24aTLAS, HuangY000O24EMRMerging, 0002LLJGYLGTH024PCBMerging, ShirafujiTT25BiasVector, JinHXS025, YoshidaNHYSSMN25} has been proposed to enable more effective knowledge consolidation. Task vectors are defined as the parameter differences between each fine-tuned model and the pre-trained model. They can be interpreted as directions that encode domain- or task-specific adaptations. Task Arithmetic~\cite{IlharcoRWSHF23TaskArith} shows that task vectors can be combined to build multi-task models, or negated to attenuate (or remove) task-specific knowledge. TIES~\cite{YadavTCRB23TIES} selects parameters with large task-induced changes and resolves sign conflicts to reduce interference.
AdaMerging~\cite{YangW00G0T24AdaMerging} extends linear task-vector composition by learning domain- or layer-wise merging weights through entropy minimization on unlabeled test samples. Although model merging has demonstrated strong effectiveness in computer vision and natural language processing, its potential for recommender systems has remained largely unexplored.
\section{Conclusion}

In this work, we addressed the fundamental limitations of existing CDSR under realistic constraints, where user interaction data cannot be shared across domains. To this end, we introduced \textit{\textbf{MergeRec}}, a novel framework that applies task vector-based model merging to a new problem setting, termed \textit{data-isolated CDSR}. MergeRec consists of three key components: (1) \textit{Merging initialization} constructs an initial merged model using training-free task vectors based merging. (2) \textit{Pseudo-user data construction} synthesizes virtual interaction sequences from domain items, allowing CF signals to be extracted without exposing sensitive user data. (3) \textit{Collaborative merging optimization} jointly optimizes a recommendation loss and a knowledge distillation loss, facilitating the transfer of domain-specific CF patterns while preserving ranking effectiveness.
Extensive experiments confirmed that MergeRec consistently outperforms existing model merging baselines and significantly improves generalization, including on unseen and data-scarce domains, highlighting the potential of model merging as a scalable, privacy-preserving solution for building universal recommender systems.

\section*{Ethical Use of Data}
This paper utilizes publicly available datasets (Amazon product review datasets) that contain no personally identifiable information and require no Institutional Ethics Review Board approval. All experiments use anonymized interaction sequences released for academic research purposes.

\begin{acks}
This work was supported by the Institute of Information \& communications Technology Planning \& Evaluation (IITP) grant and the National Research Foundation of Korea (NRF) grant funded by the Korea government (MSIT) (No. IITP-RS-2019-II190421, IITP-RS-2022-II220680, RS-2025-25442569, NRF-RS-2025-00564083, each contributing 25\% to this research).
\end{acks}

\bibliographystyle{ACM-Reference-Format}
\bibliography{citations}

@inproceedings{AkenWLG19,
  author       = {Betty van Aken and
                  Benjamin Winter and
                  Alexander L{\"{o}}ser and
                  Felix A. Gers},
  title        = {How Does {BERT} Answer Questions?: {A} Layer-Wise Analysis of Transformer
                  Representations},
  booktitle    = {CIKM},
  pages        = {1823--1832},
  year         = {2019},
}

@article{RogersKR20,
  author       = {Anna Rogers and
                  Olga Kovaleva and
                  Anna Rumshisky},
  title        = {A Primer in BERTology: What We Know About How {BERT} Works},
  journal      = {Trans. Assoc. Comput. Linguistics},
  volume       = {8},
  pages        = {842--866},
  year         = {2020},
}

@article{hou2024blair,
  title={Bridging Language and Items for Retrieval and Recommendation},
  author={Hou, Yupeng and Li, Jiacheng and He, Zhankui and Yan, An and Chen, Xiusi and McAuley, Julian},
  journal={arXiv preprint arXiv:2403.03952},
  year={2024}
}

@inproceedings{LiWLFSSM23recformer,
  author       = {Jiacheng Li and
                  Ming Wang and
                  Jin Li and
                  Jinmiao Fu and
                  Xin Shen and
                  Jingbo Shang and
                  Julian J. McAuley},
  title        = {Text Is All You Need: Learning Language Representations for Sequential
                  Recommendation},
  booktitle    = {KDD},
  pages        = {1258--1267},
  year         = {2023},
}

@inproceedings{TianCNSL22,
  author       = {Yu Tian and
                  Jianxin Chang and
                  Yanan Niu and
                  Yang Song and
                  Chenliang Li},
  title        = {When Multi-Level Meets Multi-Interest: {A} Multi-Grained Neural Model
                  for Sequential Recommendation},
  booktitle    = {SIGIR},
  pages        = {1632--1641},
  year         = {2022},
}

@inproceedings{CenZZZYT20,
  author       = {Yukuo Cen and
                  Jianwei Zhang and
                  Xu Zou and
                  Chang Zhou and
                  Hongxia Yang and
                  Jie Tang},
  title        = {Controllable Multi-Interest Framework for Recommendation},
  booktitle    = {KDD},
  pages        = {2942--2951},
  year         = {2020},
}

@article{abs-2102-04925,
  author       = {Chuhan Wu and
                  Fangzhao Wu and
                  Yang Cao and
                  Yongfeng Huang and
                  Xing Xie},
  title        = {FedGNN: Federated Graph Neural Network for Privacy-Preserving Recommendation},
  journal      = {CoRR},
  volume       = {abs/2102.04925},
  year         = {2021},
}

@incollection{YangTZCY20,
  author       = {Liu Yang and
                  Ben Tan and
                  Vincent W. Zheng and
                  Kai Chen and
                  Qiang Yang},
  title        = {Federated Recommendation Systems},
  booktitle    = {Federated Learning - Privacy and Incentive},
  volume       = {12500},
  pages        = {225--239},
  publisher    = {Springer},
  year         = {2020},
}

@inproceedings{YangPW00PW0F24,
  author       = {Ziqi Yang and
                  Zhaopeng Peng and
                  Zihui Wang and
                  Jianzhong Qi and
                  Chaochao Chen and
                  Weike Pan and
                  Chenglu Wen and
                  Cheng Wang and
                  Xiaoliang Fan},
  title        = {Federated Graph Learning for Cross-Domain Recommendation},
  booktitle    = {NeurIPS},
  year         = {2024},
}

@article{TianXCLZ24,
  author       = {Changxin Tian and
                  Yuexiang Xie and
                  Xu Chen and
                  Yaliang Li and
                  Xin Zhao},
  title        = {Privacy-preserving Cross-domain Recommendation with Federated Graph
                  Learning},
  journal      = {{ACM} Trans. Inf. Syst.},
  volume       = {42},
  number       = {5},
  pages        = {135:1--135:29},
  year         = {2024},
  url          = {https://doi.org/10.1145/3653448},
}

@article{WangSZWX24,
  author       = {Li Wang and
                  Lei Sang and
                  Quangui Zhang and
                  Qiang Wu and
                  Min Xu},
  title        = {A privacy-preserving framework with multi-modal data for cross-domain
                  recommendation},
  journal      = {Knowl. Based Syst.},
  volume       = {304},
  pages        = {112529},
  year         = {2024},
}

@article{abs-2503-23026,
  author       = {Ziang Lu and
                  Lei Guo and
                  Xu Yu and
                  Zhiyong Cheng and
                  Xiaohui Han and
                  Lei Zhu},
  title        = {Federated Semantic Learning for Privacy-preserving Cross-domain Recommendation},
  journal      = {CoRR},
  volume       = {abs/2503.23026},
  year         = {2025},
}

@inproceedings{WuLTRWX22,
  author       = {Meihan Wu and
                  Li Li and
                  Chang Tao and
                  Eric Rigall and
                  Xiaodong Wang and
                  Cheng{-}Zhong Xu},
  title        = {FedCDR: Federated Cross-Domain Recommendation for Privacy-Preserving
                  Rating Prediction},
  booktitle    = {CIKM},
  pages        = {2179--2188},
  year         = {2022},
}

@inproceedings{0002XP0024,
  author       = {Shu Chen and
                  Zitao Xu and
                  Weike Pan and
                  Qiang Yang and
                  Zhong Ming},
  title        = {A Survey on Cross-Domain Sequential Recommendation},
  booktitle    = {IJCAI},
  pages        = {7989--7998},
  year         = {2024},
}

@inproceedings{CaoCSLW22,
  author       = {Jiangxia Cao and
                  Xin Cong and
                  Jiawei Sheng and
                  Tingwen Liu and
                  Bin Wang},
  title        = {Contrastive Cross-Domain Sequential Recommendation},
  booktitle    = {CIKM},
  pages        = {138--147},
  year         = {2022},
}

@inproceedings{LinGZCNSGLJLW24,
  author       = {Guanyu Lin and
                  Chen Gao and
                  Yu Zheng and
                  Jianxin Chang and
                  Yanan Niu and
                  Yang Song and
                  Kun Gai and
                  Zhiheng Li and
                  Depeng Jin and
                  Yong Li and
                  Meng Wang},
  title        = {Mixed Attention Network for Cross-domain Sequential Recommendation},
  booktitle    = {WSDM},
  pages        = {405--413},
  year         = {2024},
}

@article{MaRCRZLMR22,
  author       = {Muyang Ma and
                  Pengjie Ren and
                  Zhumin Chen and
                  Zhaochun Ren and
                  Lifan Zhao and
                  Peiyu Liu and
                  Jun Ma and
                  Maarten de Rijke},
  title        = {Mixed Information Flow for Cross-Domain Sequential Recommendations},
  journal      = {{ACM} Trans. Knowl. Discov. Data},
  volume       = {16},
  number       = {4},
  pages        = {64:1--64:32},
  year         = {2022},
}

@inproceedings{00020W0Z0LH025,
  author       = {Qidong Liu and
                  Xiangyu Zhao and
                  Yejing Wang and
                  Zijian Zhang and
                  Howard Zhong and
                  Chong Chen and
                  Xiang Li and
                  Wei Huang and
                  Feng Tian},
  title        = {Bridge the Domains: Large Language Models Enhanced Cross-domain Sequential
                  Recommendation},
  booktitle    = {SIGIR},
  pages        = {1582--1592},
  year         = {2025},
}

@inproceedings{ZhuC0LZ19,
  author       = {Feng Zhu and
                  Chaochao Chen and
                  Yan Wang and
                  Guanfeng Liu and
                  Xiaolin Zheng},
  title        = {{DTCDR:} {A} Framework for Dual-Target Cross-Domain Recommendation},
  booktitle    = {CIKM},
  pages        = {1533--1542},
  year         = {2019},
}

@article{WangYZZC25,
  author       = {Hao Wang and
                  Mingjia Yin and
                  Luankang Zhang and
                  Sirui Zhao and
                  Enhong Chen},
  title        = {{MF-GSLAE:} {A} Multi-Factor User Representation Pre-Training Framework
                  for Dual-Target Cross-Domain Recommendation},
  journal      = {{ACM} Trans. Inf. Syst.},
  volume       = {43},
  number       = {2},
  pages        = {30:1--30:28},
  year         = {2025},

}

@inproceedings{ParkKCHYCLRYCC23,
  author       = {Chung Park and
                  Taesan Kim and
                  Taekyoon Choi and
                  Junui Hong and
                  Yelim Yu and
                  Mincheol Cho and
                  Kyunam Lee and
                  Sungil Ryu and
                  Hyungjun Yoon and
                  Minsung Choi and
                  Jaegul Choo},
  title        = {Cracking the Code of Negative Transfer: {A} Cooperative Game Theoretic
                  Approach for Cross-Domain Sequential Recommendation},
  booktitle    = {CIKM},
  pages        = {2024--2033},
  year         = {2023},
}

@inproceedings{ParkKYHYCCC24,
  author       = {Chung Park and
                  Taesan Kim and
                  Hyungjun Yoon and
                  Junui Hong and
                  Yelim Yu and
                  Mincheol Cho and
                  Minsung Choi and
                  Jaegul Choo},
  title        = {Pacer and Runner: Cooperative Learning Framework between Single- and
                  Cross-Domain Sequential Recommendation},
  booktitle    = {SIGIR},
  pages        = {2071--2080},
  year         = {2024},
}

@inproceedings{IlharcoWGSHKFS22AvgMer1,
  author       = {Gabriel Ilharco and
                  Mitchell Wortsman and
                  Samir Yitzhak Gadre and
                  Shuran Song and
                  Hannaneh Hajishirzi and
                  Simon Kornblith and
                  Ali Farhadi and
                  Ludwig Schmidt},
  title        = {Patching open-vocabulary models by interpolating weights},
  booktitle    = {NeurIPS},
  year         = {2022},
}

@inproceedings{WortsmanIKLKRLH22AvgMer2,
  author       = {Mitchell Wortsman and
                  Gabriel Ilharco and
                  Jong Wook Kim and
                  Mike Li and
                  Simon Kornblith and
                  Rebecca Roelofs and
                  Raphael Gontijo Lopes and
                  Hannaneh Hajishirzi and
                  Ali Farhadi and
                  Hongseok Namkoong and
                  Ludwig Schmidt},
  title        = {Robust fine-tuning of zero-shot models},
  booktitle    = {CVPR},
  pages        = {7949--7961},
  year         = {2022},
}

@inproceedings{IlharcoRWSHF23TaskArith,
  author       = {Gabriel Ilharco and
                  Marco T{\'{u}}lio Ribeiro and
                  Mitchell Wortsman and
                  Ludwig Schmidt and
                  Hannaneh Hajishirzi and
                  Ali Farhadi},
  title        = {Editing models with task arithmetic},
  booktitle    = {ICLR},
  year         = {2023},
}

@inproceedings{YadavTCRB23TIES,
  author       = {Prateek Yadav and
                  Derek Tam and
                  Leshem Choshen and
                  Colin A. Raffel and
                  Mohit Bansal},
  title        = {TIES-Merging: Resolving Interference When Merging Models},
  booktitle    = {NeurIPS},
  year         = {2023},
}

@inproceedings{YangW00G0T24AdaMerging,
  author       = {Enneng Yang and
                  Zhenyi Wang and
                  Li Shen and
                  Shiwei Liu and
                  Guibing Guo and
                  Xingwei Wang and
                  Dacheng Tao},
  title        = {AdaMerging: Adaptive Model Merging for Multi-Task Learning},
  booktitle    = {ICLR},
  year         = {2024},
}

@inproceedings{JimenezFF23,
  author       = {Guillermo Ortiz{-}Jim{\'{e}}nez and
                  Alessandro Favero and
                  Pascal Frossard},
  title        = {Task Arithmetic in the Tangent Space: Improved Editing of Pre-Trained
                  Models},
  booktitle    = {NeurIPS 2023},
  year         = {2023},
}

@inproceedings{WangDOFF24Consensus,
  author       = {Ke Wang and
                  Nikolaos Dimitriadis and
                  Guillermo Ortiz{-}Jim{\'{e}}nez and
                  Fran{\c{c}}ois Fleuret and
                  Pascal Frossard},
  title        = {Localizing Task Information for Improved Model Merging and Compression},
  booktitle    = {ICML},
  year         = {2024},
}

@inproceedings{ZhangAOHA24aTLAS,
  author       = {Frederic Z. Zhang and
                  Paul Albert and
                  Cristian Rodriguez Opazo and
                  Anton van den Hengel and
                  Ehsan Abbasnejad},
  title        = {Knowledge Composition using Task Vectors with Learned Anisotropic
                  Scaling},
  booktitle    = {NeurIPS 2024},
  year         = {2024},
}

@inproceedings{HuangY000O24EMRMerging,
  author       = {Chenyu Huang and
                  Peng Ye and
                  Tao Chen and
                  Tong He and
                  Xiangyu Yue and
                  Wanli Ouyang},
  title        = {EMR-Merging: Tuning-Free High-Performance Model Merging},
  booktitle    = {NeurIPS},
  year         = {2024},
}

@inproceedings{0002LLJGYLGTH024PCBMerging,
  author       = {Guodong Du and
                  Junlin Lee and
                  Jing Li and
                  Runhua Jiang and
                  Yifei Guo and
                  Shuyang Yu and
                  Hanting Liu and
                  Sim Kuan Goh and
                  Ho{-}Kin Tang and
                  Daojing He and
                  Min Zhang},
  title        = {Parameter Competition Balancing for Model Merging},
  booktitle    = {NeurIPS},
  year         = {2024},
}

@inproceedings{ShirafujiTT25BiasVector,
  author       = {Daiki Shirafuji and
                  Makoto Takenaka and
                  Shinya Taguchi},
  title        = {Bias Vector: Mitigating Biases in Language Models with Task Arithmetic
                  Approach},
  booktitle    = {COLING},
  pages        = {2799--2813},
  year         = {2025},
}

@inproceedings{JinHXS025,
  author       = {Ruochen Jin and
                  Bojian Hou and
                  Jiancong Xiao and
                  Weijie J. Su and
                  Li Shen},
  title        = {Fine-Tuning Attention Modules Only: Enhancing Weight Disentanglement
                  in Task Arithmetic},
  booktitle    = {ICLR},
  year         = {2025},
}

@inproceedings{YoshidaNHYSSMN25,
  author       = {Kotaro Yoshida and
                  Yuji Naraki and
                  Takafumi Horie and
                  Ryosuke Yamaki and
                  Ryotaro Shimizu and
                  Yuki Saito and
                  Julian J. McAuley and
                  Hiroki Naganuma},
  title        = {Mastering Task Arithmetic: {\(\tau\)}Jp as a Key Indicator for Weight
                  Disentanglement},
  booktitle    = {ICLR},
  year         = {2025},
}

@inproceedings{MatenaR22Fisher,
  author       = {Michael Matena and
                  Colin Raffel},
  title        = {Merging Models with Fisher-Weighted Averaging},
  booktitle    = {NeurIPS},
  year         = {2022},
}

@inproceedings{XuYWWSS24MuDSC,
  author       = {Zhengqi Xu and
                  Ke Yuan and
                  Huiqiong Wang and
                  Yong Wang and
                  Mingli Song and
                  Jie Song},
  title        = {Training-Free Pretrained Model Merging},
  booktitle    = {CVPR},
  pages        = {5915--5925},
  year         = {2024},
}

@inproceedings{StoicaBBRHH24ZipIt,
  author       = {George Stoica and
                  Daniel Bolya and
                  Jakob Bjorner and
                  Pratik Ramesh and
                  Taylor Hearn and
                  Judy Hoffman},
  title        = {ZipIt! Merging Models from Different Tasks without Training},
  booktitle    = {ICLR},
  year         = {2024},
}

@inproceedings{Yang0WG00T24Surgery,
  author       = {Enneng Yang and
                  Li Shen and
                  Zhenyi Wang and
                  Guibing Guo and
                  Xiaojun Chen and
                  Xingwei Wang and
                  Dacheng Tao},
  title        = {Representation Surgery for Multi-Task Model Merging},
  booktitle    = {ICML},
  year         = {2024},
}

@inproceedings{LuF0QC024TwinMerging,
  author       = {Zhenyi Lu and
                  Chenghao Fan and
                  Wei Wei and
                  Xiaoye Qu and
                  Dangyang Chen and
                  Yu Cheng},
  title        = {Twin-Merging: Dynamic Integration of Modular Expertise in Model Merging},
  booktitle    = {NeurIPS 2024},
  year         = {2024},
}

@inproceedings{GargiuloCBSSR25TSV,
  author       = {Antonio Andrea Gargiulo and
                  Donato Crisostomi and
                  Maria Sofia Bucarelli and
                  Simone Scardapane and
                  Fabrizio Silvestri and
                  Emanuele Rodol{\`{a}}},
  title        = {Task Singular Vectors: Reducing Task Interference in Model Merging},
  booktitle    = {CVPR},
  pages        = {18695--18705},
  year         = {2025},
}

@inproceedings{KangM18SASRec,
  author    = {Wang{-}Cheng Kang and
               Julian J. McAuley},
  title     = {Self-Attentive Sequential Recommendation},
  booktitle = {ICDM},
  pages     = {197--206},
  year      = {2018},
}

@inproceedings{SunLWPLOJ19BERT4Rec,
  author    = {Fei Sun and
               Jun Liu and
               Jian Wu and
               Changhua Pei and
               Xiao Lin and
               Wenwu Ou and
               Peng Jiang},
  title     = {BERT4Rec: Sequential Recommendation with Bidirectional Encoder Representations
               from Transformer},
  booktitle = {CIKM},
  pages     = {1441--1450},
  year      = {2019},
}

@inproceedings{HidasiKBT15GRU4Rec,
  author    = {Bal{\'{a}}zs Hidasi and
               Alexandros Karatzoglou and
               Linas Baltrunas and
               Domonkos Tikk},
  title     = {Session-based Recommendations with Recurrent Neural Networks},
  booktitle = {ICLR},
  year      = {2016},
}

@inproceedings{ZhangYYLFZC022Re4,
  author    = {Shengyu Zhang and
               Lingxiao Yang and
               Dong Yao and
               Yujie Lu and
               Fuli Feng and
               Zhou Zhao and
               Tat{-}Seng Chua and
               Fei Wu},
  title     = {Re4: Learning to Re-contrast, Re-attend, Re-construct for Multi-interest
               Recommendation},
  booktitle = {WWW},
  pages     = {2216--2226},
  year      = {2022},
}

@inproceedings{LiZL0WG22MLP4Rec,
  author       = {Muyang Li and
                  Xiangyu Zhao and
                  Chuan Lyu and
                  Minghao Zhao and
                  Runze Wu and
                  Ruocheng Guo},
  title        = {MLP4Rec: {A} Pure {MLP} Architecture for Sequential Recommendations},
  booktitle    = {IJCAI},
  pages        = {2138--2144},
  publisher    = {ijcai.org},
  year         = {2022},
}

@inproceedings{LiZZW0WG23AutoMLP,
  author       = {Muyang Li and
                  Zijian Zhang and
                  Xiangyu Zhao and
                  Wanyu Wang and
                  Minghao Zhao and
                  Runze Wu and
                  Ruocheng Guo},
  title        = {AutoMLP: Automated {MLP} for Sequential Recommendations},
  booktitle    = {WWW},
  pages        = {1190--1198},
  publisher    = {{ACM}},
  year         = {2023},
}

@article{MoonJCCSL23,
  author       = {Jaewan Moon and
                  Yoonki Jeong and
                  Dong{-}Kyu Chae and
                  Jaeho Choi and
                  Hyunjung Shim and
                  Jongwuk Lee},
  title        = {CoMix: Collaborative filtering with mixup for implicit datasets},
  journal      = {Inf. Sci.},
  volume       = {628},
  pages        = {254--268},
  year         = {2023},
}

@inproceedings{MoonKL23,
  author       = {Jaewan Moon and
                  Hye{-}young Kim and
                  Jongwuk Lee},
  title        = {It's Enough: Relaxing Diagonal Constraints in Linear Autoencoders
                  for Recommendation},
  booktitle    = {SIGIR},
  pages        = {1639--1648},
  year         = {2023},
}

@inproceedings{Moon0L25,
  author       = {Jaewan Moon and
                  Seongmin Park and
                  Jongwuk Lee},
  title        = {LLM-Enhanced Linear Autoencoders for Recommendation},
  booktitle    = {CIKM},
  pages        = {5036--5040},
  year         = {2025},
}

@inproceedings{ParkYLPL23MAWU,
  author       = {Seongmin Park and
                  Mincheol Yoon and
                  Jae{-}woong Lee and
                  Hogun Park and
                  Jongwuk Lee},
  title        = {Toward a Better Understanding of Loss Functions for Collaborative
                  Filtering},
  booktitle    = {CIKM},
  pages        = {2034--2043},
  year         = {2023},
}

@inproceedings{park2025dan,
  title        = {Why is Normalization Necessary for Linear Recommenders?},
  author       = {Seongmin Park and
                  Mincheol Yoon and
                  Hye-young Kim and
                  Jongwuk Lee},
  booktitle    = {SIGIR},
  pages        = {2142--2151},
  year         = {2025},
}

\newpage
\appendix

\section{Dataset Statistics}\label{app:statistics}
\begin{table}[!t]
\caption{Dataset statistics including the number of users, items, interactions, and density.} \label{tab:data_statistics}

\centering
\begin{tabular}{l|rrrr}
\toprule
\textbf{Dataset} & \textbf{\# Users} & \textbf{\# Items} & \textbf{\# Inter.} & \textbf{Density} \\
\midrule
Arts   & 56,210  & 22,855  & 492,492  & 0.04\% \\
Beauty & 22,363  & 12,101  & 198,502  & 0.07\% \\
Inst.  & 27,530  & 10,611  & 231,312  & 0.08\% \\
Office & 101,499 & 27,932  & 798,912  & 0.03\% \\
Pantry & 14,180  & 4,968   & 137,769  & 0.20\% \\
Sci.   & 11,041  & 5,327   & 76,896   & 0.13\% \\
Sports & 35,598  & 18,357  & 296,337  & 0.05\% \\
Toys   & 19,412  & 11,924  & 167,597  & 0.07\% \\
\bottomrule
\end{tabular}
\end{table}

Table~\ref{tab:data_statistics} summarizes the statistics of each domain, including the number of users, items, interactions, and dataset density. The density is calculated as $\frac{\#~\text{Interactions}}{\#~\text{Users} \times \#~\text{Items}}$.

\section{Implementation Details}\label{app:implementation}
All methods, including MergeRec and baselines, are implemented in PyTorch. For RecFormer-base~\cite{LiWLFSSM23recformer}, we use the official pre-trained checkpoint\footnote{\url{https://github.com/AaronHeee/RecFormer}}, while for RecFormer-large, we pre-train the model following the protocol described in the original paper.
For BLaIR-base\footnote{\url{https://huggingface.co/hyp1231/blair-roberta-base}} and BLaIR-large\footnote{\url{https://huggingface.co/hyp1231/blair-roberta-large}}~\cite{hou2024blair}, we use the official pre-trained checkpoints available on HuggingFace. Fine-tuning is performed with in-batch negative sampling and a batch size of 64. For merging baselines, we adopt their hyperparameter configurations for the validation-less setting, \ie, no training, validation, or test data are used. Specifically, we set $w_1 = w_2 = \dots = w_K = 0.4$ for Task Arithmetic~\cite{IlharcoRWSHF23TaskArith}, and $w=1$ for TIES~\cite{YadavTCRB23TIES}. We use the top 20\% of the parameters for TIES. For AdaMerging~\cite{YangW00G0T24AdaMerging} and MergeRec, all coefficients are initialized to $0.2$ and optimized for 500 steps using the Adam optimizer with a learning rate of 0.001 and a batch size of 16. For MergeRec, we set $\lambda = 1{,}000$ to balance the scale of the two loss functions.
All reported performance metrics represent averages computed across five random seeds. For the significance test, we assume that deterministic merging approaches (\ie, Weight Averaging, Task Arithmetic, and TIES) have identical performance values across all five runs.

\section{Overall Performance on Other Metrics}\label{app:results}
Table~\ref{tab:overall_ndcg} shows the normalized N@10 of MergeRec and seven baseline methods across eight datasets on four backbone models, with the performance of the fine-tuned model trained on each dataset set to 100\%.
In addition, we report the unnormalized R@10 and N@10 results in Table~\ref{tab:overall_recall_abs} and Table~\ref{tab:overall_ndcg_abs}, respectively.
We observe similar trends for N@10 (Table~\ref{tab:overall_ndcg}) as for R@10 (Table~\ref{tab:overall}).
\textbf{(i)} MergeRec achieves the best performance on average across all datasets and backbone models. This shows that MergeRec effectively transfers domain-specific knowledge and improves model discrimination, whereas AdaMerging is less effective across the board.
\textbf{(ii)} Merging methods like MergeRec, Weight Averaging, and TIES consistently outperform joint learning, demonstrating that parameter merging can efficiently transfer knowledge across domains and reduce computational cost without cross-domain training data.
\textbf{(iii)} Cross-domain merging is especially helpful for domains with limited data, such as Scientific, where MergeRec shows notable improvements over fine-tuned models due to more effective knowledge transfer.

\section{Domain-Specific Weight Dynamics on Other Backbones} \label{domain_weights_other_backbones}

We further analyze the evolution of domain-specific weights $\mathbf{w}$ on the remaining three backbone models: RecFormer-large and BLaIR-base/large. Consistent with our observations for RecFormer-base, we find that domains with larger data scales (\eg, \textit{Arts} and \textit{Office}) tend to converge to higher weight coefficients. This trend suggests that these models also prioritize capturing more complex collaborative patterns in large-scale domains, thereby allocating more representational capacity to preserve domain-specific knowledge.
In contrast to this general trend, the merging weight for the Pantry dataset in the BLaIR model converges to negative values.
This behavior may be attributed to the absence of category information in the Pantry dataset, which differs from other domains and can lead to misalignment during model merging.

\begin{figure} [t]
\centering
(a) Recformer-large
\includegraphics[width=0.9\linewidth]{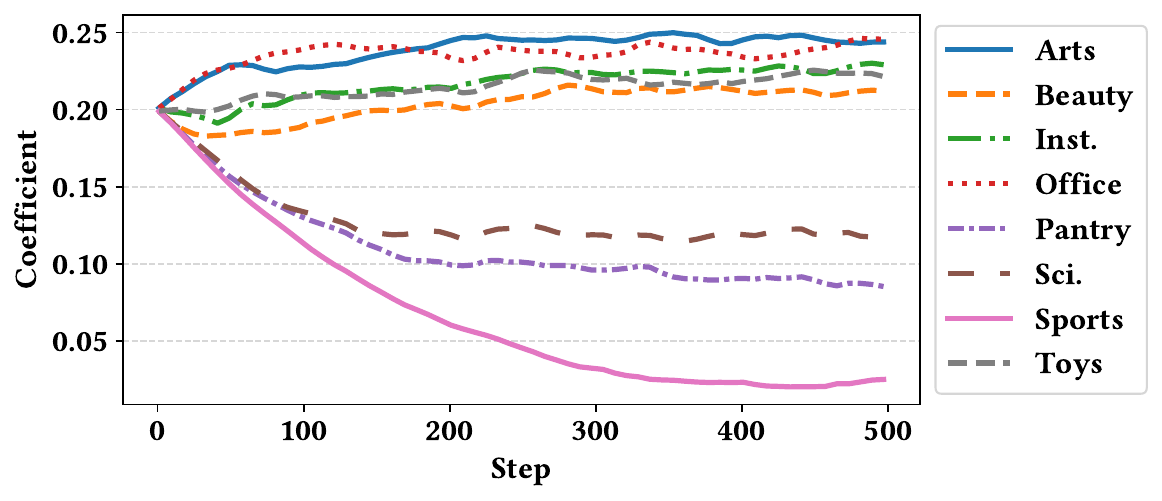}

(b) BLaIR-base
\includegraphics[width=0.9\linewidth]{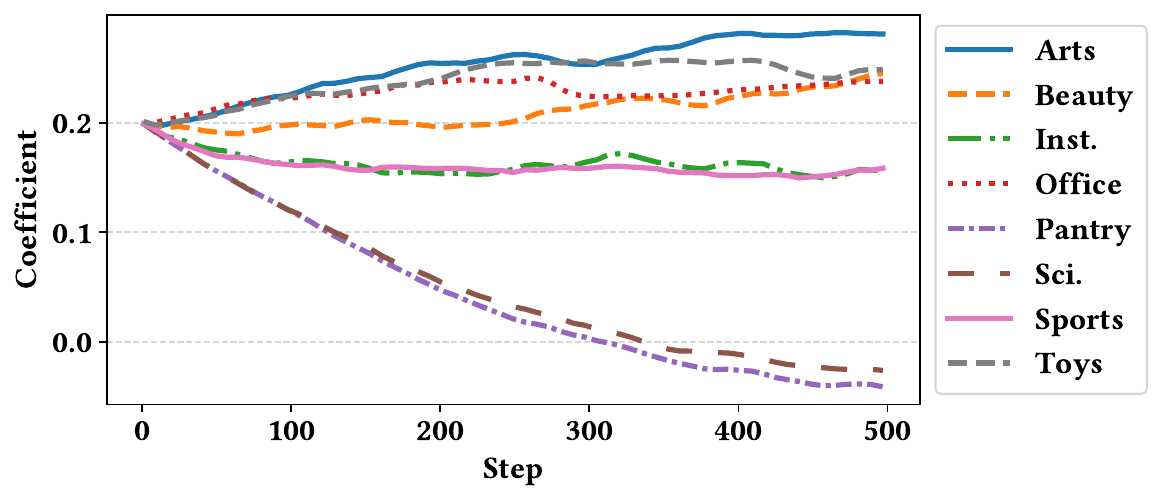}

(c) BLaIR-large
\includegraphics[width=0.9\linewidth]{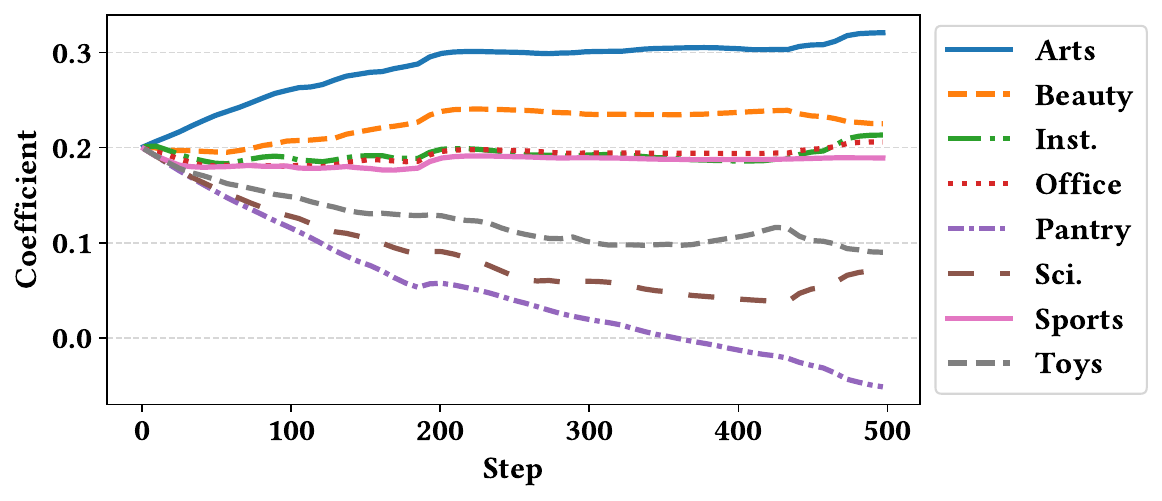}

\vspace{-3mm}
\caption{Domain-specific coefficients dynamics over training steps on RecFormer-large and BLaIR-base/large backbones.}
\label{fig:model_coefficients_over_training_app}
\vspace{-2mm}
\end{figure}

\begin{table*}[t] \small
\caption{Performance comparison with six baseline methods on four backbone models, \ie, RecFormer-base/large~\cite{LiWLFSSM23recformer} and BLaIR-base/large~\cite{hou2024blair}. We report normalized NDCG@10 performance (\%) where the fine-tuned model's performance is 100\%. The best results are marked in \textbf{bold}, and the second-best results are shown as \underline{underlined}. `*’ indicates the statistically significant gain of MergeRec over the best baseline model (p < 0.02 for one-tailed t-test).}

\centering
\label{tab:overall_ndcg}

\begin{tabular}{c|l|>{\columncolor{gray!10}}rrrrrrrrr}
\toprule
\multicolumn{1}{c|}{\textbf{Backbone}} & 
\multicolumn{1}{c|}{\textbf{Method}} & 
\multicolumn{1}{c}{\textbf{Avg.}} & 
\multicolumn{1}{c}{Arts} & 
\multicolumn{1}{c}{Beauty} & 
\multicolumn{1}{c}{Inst.} & 
\multicolumn{1}{c}{Office} & 
\multicolumn{1}{c}{Pantry} & 
\multicolumn{1}{c}{Sci.} & 
\multicolumn{1}{c}{Sports} & 
\multicolumn{1}{c}{Toys} \\
\midrule

\multirow{9.8}{*}{RecFormer-base}
& Zero-shot & 67.52 & 63.59 & 61.57 & 55.88 & 52.63 & 69.33 & 92.79 & 70.75 & 81.44 \\
& Joint Learning & 73.05 & 67.41 & 75.48 & 76.13 & 66.69 & 82.75 & 72.12 & 69.55 & \textbf{87.62} \\
\cmidrule{2-11}
& Weight Averaging & 83.15 & 82.88 & 76.42 & 75.67 & 70.29 & 92.37 & 99.68 & \textbf{82.70} & 86.20 \\
& Task Arithmetic & 84.08 & 86.47 & 75.99 & 76.05 & 76.37 & 91.85 & 99.52 & 78.08 & 75.53 \\
& TIES & \uline{85.60} & \uline{87.16} & \textbf{79.61} & 76.52 & \textbf{81.25} & 91.95 & 97.59 & 80.72 & 80.11 \\
& AdaMerging (Domain-wise) & 71.17 & 75.47 & 55.19 & 60.35 & 56.45 & 71.09 & 90.67 & 67.83 & \uline{87.15} \\
& AdaMerging (Layer-wise) & 61.61 & 60.94 & 32.01 & 52.35 & 51.36 & 64.14 & 82.60 & 64.32 & 79.22 \\
\cmidrule{2-11}
& MergeRec (Domain-wise) & \textbf{86.07}* & \textbf{87.31} & 76.54 & \textbf{79.14}* & \uline{76.87} & \textbf{94.50}* & \uline{100.48}* & 80.36 & 84.97 \\
& MergeRec (Layer-wise) & 85.37 & 86.18 & \uline{77.32} & \uline{77.77}* & 75.39 & \uline{93.82}* & \textbf{101.00}* & \uline{81.55} & 83.87 \\
\midrule

\multirow{9.8}{*}{RecFormer-large}
& Zero-shot & 53.55 & 65.05 & 50.33 & 34.05 & 31.26 & 40.96 & 82.63 & 61.27 & 63.58 \\
& Joint Learning & 75.76 & 72.09 & 75.48 & 66.50 & 66.67 & 86.94 & 87.67 & 85.34 & 78.60 \\
\cmidrule{2-11}
& Weight Averaging & 85.84 & 86.37 & 81.64 & 76.44 & 74.60 & \uline{94.63} & 98.70 & \textbf{91.35} & 88.87 \\
& Task Arithmetic & 85.70 & 93.60 & 80.00 & 77.58 & 77.38 & 91.60 & 94.56 & 81.75 & 77.75 \\
& TIES & 87.64 & \uline{94.09} & 81.66 & \uline{80.60} & \textbf{82.07} & 92.36 & 96.49 & 83.51 & 76.81 \\
& AdaMerging (Domain-wise) & 67.29 & 77.00 & 56.84 & 61.87 & 56.92 & 55.66 & 82.95 & 56.31 & 69.72 \\
& AdaMerging (Layer-wise) & 64.96 & 71.01 & 51.70 & 53.68 & 54.66 & 62.19 & 81.38 & 69.86 & 69.99 \\
\cmidrule{2-11}
& MergeRec (Domain-wise) & \textbf{89.40}* & \textbf{96.27}* & \uline{84.65}* & \textbf{80.68} & \uline{78.21} & 94.07 & \uline{98.97} & 88.13 & \uline{90.48}* \\
& MergeRec (Layer-wise) & \uline{88.62}* & 93.40 & \textbf{85.08}* & 78.27 & 77.58 & \textbf{94.80} & \textbf{99.58} & \uline{89.95} & \textbf{90.66}* \\
\midrule

\multirow{9.8}{*}{BLaIR-base}
& Zero-shot & 33.06 & 32.86 & 33.22 & 23.98 & 20.80 & 40.87 & 45.76 & 34.05 & 40.78 \\
& Joint Learning & \uline{78.48} & 79.94 & \textbf{97.16} & \textbf{79.19} & 67.59 & \textbf{88.03} & 68.00 & \textbf{92.07} & \textbf{92.08} \\
\cmidrule{2-11}
& Weight Averaging & 78.13 & 78.59 & \uline{85.53} & \uline{67.01} & 64.70 & \uline{85.65} & \textbf{94.62} & \uline{82.03} & 75.85 \\
& Task Arithmetic & 55.37 & 50.11 & 51.82 & 55.54 & 52.30 & 58.80 & 66.56 & 49.09 & 51.86 \\
& TIES & 74.72 & 72.17 & 73.04 & 65.61 & \textbf{78.62} & 71.78 & 82.47 & 81.81 & 73.44 \\
& AdaMerging (Domain-wise) & 50.28 & 51.53 & 49.02 & 41.46 & 35.38 & 43.17 & 66.06 & 68.61 & 69.18 \\
& AdaMerging (Layer-wise) & 57.91 & 56.90 & 56.93 & 53.60 & 44.83 & 49.51 & 77.66 & 69.20 & 65.29 \\
\cmidrule{2-11}
& MergeRec (Domain-wise) & 78.30 & \textbf{81.33}* & 81.78 & 65.59 & \uline{70.51} & 78.22 & 92.14 & 79.46 & \uline{78.36} \\
& MergeRec (Layer-wise) & \textbf{79.18} & \uline{81.23}* & 84.30 & 66.94 & 70.26 & 81.88 & \uline{93.35} & 81.73 & 77.47 \\
\midrule

\multirow{9.8}{*}{BLaIR-large}
& Zero-shot & 26.85 & 27.51 & 27.83 & 18.08 & 14.22 & 41.96 & 32.15 & 28.46 & 38.78 \\
& Joint Learning & 81.67 & 81.00 & 80.58 & \textbf{85.57} & \uline{75.84} & \textbf{92.12} & 71.85 & 85.25 & \textbf{97.07} \\
\cmidrule{2-11}
& Weight Averaging & 81.50 & 83.19 & 83.57 & 71.84 & 67.74 & 90.68 & 98.42 & 84.70 & 76.31 \\
& Task Arithmetic & 73.00 & 73.33 & 77.37 & 57.23 & 66.13 & 81.75 & 89.62 & 75.40 & 65.31 \\
& TIES & \uline{85.74} & 85.71 & \uline{87.92} & 76.38 & \textbf{83.56} & 89.96 & 97.93 & \textbf{90.85} & 72.04 \\
& AdaMerging (Domain-wise) & 60.21 & 68.38 & 58.18 & 57.61 & 42.43 & 56.16 & 72.01 & 75.84 & 62.50 \\
& AdaMerging (Layer-wise) & 70.67 & 74.10 & 69.21 & 66.94 & 54.57 & 80.67 & 79.05 & 70.79 & 77.29 \\
\cmidrule{2-11}
& MergeRec (Domain-wise) & 84.46 & \uline{86.73}* & 85.69 & 78.68 & 72.36 & 87.45 & \uline{100.83}* & 86.61 & 77.22 \\
& MergeRec (Layer-wise) & \textbf{86.65}* & \textbf{87.64}* & \textbf{88.94}* & \uline{79.21} & 74.91 & \uline{91.43} & \textbf{104.19}* & \uline{87.27} & \uline{79.87} \\

\bottomrule
\end{tabular}
\vspace{-2mm}
\end{table*}

\begin{table*}[t] \small
\caption{Performance comparison with seven baseline methods on four backbone models, \ie, RecFormer-base/large~\cite{LiWLFSSM23recformer} and BLaIR-base/large~\cite{hou2024blair}. We report absolute Recall@10 performance. The best results, excluding the fine-tuned model, are marked in \textbf{bold}, and the second-best results are shown as \underline{underlined}. `*’ indicates the statistically significant gain of MergeRec over the best baseline model (p < 0.02 for one-tailed t-test).}

\centering
\label{tab:overall_recall_abs}

\begin{tabular}{c|l|>{\columncolor{gray!10}}rrrrrrrrr}
\toprule
\multicolumn{1}{c|}{\textbf{Backbone}} & 
\multicolumn{1}{c|}{\textbf{Method}} & 
\multicolumn{1}{c}{\textbf{Avg.}} & 
\multicolumn{1}{c}{Arts} & 
\multicolumn{1}{c}{Beauty} & 
\multicolumn{1}{c}{Inst.} & 
\multicolumn{1}{c}{Office} & 
\multicolumn{1}{c}{Pantry} & 
\multicolumn{1}{c}{Sci.} & 
\multicolumn{1}{c}{Sports} & 
\multicolumn{1}{c}{Toys} \\
\midrule

\multirow{9.8}{*}{RecFormer-base}
& Zero-shot & 0.0767 & 0.1192 & 0.0445 & 0.0717 & 0.0850 & 0.0664 & 0.1309 & 0.0255 & 0.0706 \\
& Fine-tune & 0.1017 & 0.1567 & 0.0699 & 0.0994 & 0.1379 & 0.0895 & 0.1409 & 0.0358 & 0.0831 \\
& Joint Learning & 0.0815 & 0.1212 & 0.0556 & 0.0828 & 0.1012 & 0.0764 & 0.1127 & 0.0249 & 0.0773 \\
\cmidrule{2-11}
& Weight Averaging & 0.0913 & 0.1429 & 0.0580 & 0.0903 & 0.1067 & 0.0838 & 0.1405 & \textbf{0.0310} & 0.0773 \\
& Task Arithmetic & 0.0904 & 0.1434 & 0.0569 & 0.0876 & 0.1156 & 0.0829 & 0.1394 & 0.0291 & 0.0685 \\
& TIES & 0.0926 & 0.1462 & \textbf{0.0598} & 0.0900 & \textbf{0.1217} & 0.0830 & 0.1377 & \uline{0.0306} & 0.0717 \\
& AdaMerging (Domain-wise) & 0.0802 & 0.1367 & 0.0410 & 0.0754 & 0.0893 & 0.0671 & 0.1305 & 0.0243 & \uline{0.0774} \\
& AdaMerging (Layer-wise) & 0.0685 & 0.1053 & 0.0230 & 0.0670 & 0.0807 & 0.0588 & 0.1192 & 0.0233 & 0.0706 \\
\cmidrule{2-11}
& MergeRec (Domain-wise) & \textbf{0.0939}* & \textbf{0.1507}* & 0.0585 & \uline{0.0904} & \uline{0.1162} & \textbf{0.0855}* & \uline{0.1420}* & 0.0301 & \textbf{0.0776}* \\
& MergeRec (Layer-wise) & \uline{0.0936}* & \uline{0.1496}* & \uline{0.0593} & \textbf{0.0906}* & 0.1145 & \uline{0.0849}* & \textbf{0.1430}* & 0.0304 & 0.0766 \\
\midrule

\multirow{9.8}{*}{RecFormer-large}
& Zero-shot & 0.0613 & 0.1138 & 0.0374 & 0.0470 & 0.0545 & 0.0415 & 0.1200 & 0.0205 & 0.0557 \\
& Fine-tune & 0.1038 & 0.1568 & 0.0721 & 0.1021 & 0.1388 & 0.0935 & 0.1463 & 0.0327 & 0.0877 \\
& Joint Learning & 0.0869 & 0.1304 & 0.0573 & 0.0815 & 0.1022 & 0.0844 & 0.1360 & 0.0299 & 0.0733 \\
\cmidrule{2-11}
& Weight Averaging & 0.0947 & 0.1447 & 0.0628 & 0.0908 & 0.1098 & \textbf{0.0902} & \textbf{0.1439} & \textbf{0.0323} & 0.0827 \\
& Task Arithmetic & 0.0913 & 0.1432 & 0.0601 & 0.0901 & 0.1149 & 0.0861 & 0.1360 & 0.0287 & 0.0711 \\
& TIES & 0.0933 & 0.1458 & 0.0618 & 0.0922 & \textbf{0.1224} & 0.0869 & 0.1377 & 0.0291 & 0.0708 \\
& AdaMerging (Domain-wise) & 0.0753 & 0.1307 & 0.0429 & 0.0792 & 0.0885 & 0.0574 & 0.1228 & 0.0190 & 0.0620 \\
& AdaMerging (Layer-wise) & 0.0735 & 0.1248 & 0.0377 & 0.0697 & 0.0842 & 0.0628 & 0.1216 & 0.0239 & 0.0631 \\
\cmidrule{2-11}
& MergeRec (Domain-wise) & \textbf{0.0965}* & \textbf{0.1492}* & \textbf{0.0650}* & \textbf{0.0937}* & \uline{0.1157} & 0.0900 & 0.1425 & 0.0310 & \textbf{0.0847}* \\
& MergeRec (Layer-wise) & \uline{0.0960}* & \uline{0.1477}* & \uline{0.0647}* & \uline{0.0922} & 0.1141 & \uline{0.0900} & \uline{0.1432} & \uline{0.0316} & \uline{0.0843}* \\
\midrule

\multirow{9.8}{*}{BLaIR-base}
& Zero-shot & 0.0409 & 0.0704 & 0.0224 & 0.0309 & 0.0376 & 0.0431 & 0.0759 & 0.0110 & 0.0355 \\
& Fine-tune & 0.0995 & 0.1535 & 0.0631 & 0.0980 & 0.1356 & 0.0913 & 0.1382 & 0.0320 & 0.0840 \\
& Joint Learning & 0.0832 & 0.1272 & \textbf{0.0614} & \textbf{0.0819} & 0.0991 & \uline{0.0833} & 0.1034 & \textbf{0.0297} & \textbf{0.0797} \\
\cmidrule{2-11}
& Weight Averaging & \uline{0.0874} & 0.1412 & \uline{0.0593} & 0.0768 & 0.1032 & \textbf{0.0834} & \textbf{0.1381} & \uline{0.0286} & 0.0687 \\
& Task Arithmetic & 0.0612 & 0.0876 & 0.0337 & 0.0653 & 0.0835 & 0.0570 & 0.1013 & 0.0158 & 0.0452 \\
& TIES & 0.0825 & 0.1295 & 0.0492 & 0.0767 & \textbf{0.1188} & 0.0705 & 0.1223 & 0.0279 & 0.0650 \\
& AdaMerging (Domain-wise) & 0.0603 & 0.1043 & 0.0329 & 0.0523 & 0.0608 & 0.0464 & 0.1008 & 0.0233 & 0.0614 \\
& AdaMerging (Layer-wise) & 0.0681 & 0.1122 & 0.0386 & 0.0675 & 0.0750 & 0.0525 & 0.1160 & 0.0238 & 0.0593 \\
\cmidrule{2-11}
& MergeRec (Domain-wise) & 0.0869 & \textbf{0.1449}* & 0.0564 & 0.0762 & \uline{0.1104} & 0.0778 & 0.1320 & 0.0272 & \uline{0.0706} \\
& MergeRec (Layer-wise) & \textbf{0.0875} & \uline{0.1436}* & 0.0579 & \uline{0.0774} & 0.1100 & 0.0802 & \uline{0.1334} & 0.0280 & 0.0698 \\
\midrule

\multirow{9.8}{*}{BLaIR-large}
& Zero-shot & 0.0333 & 0.0589 & 0.0205 & 0.0246 & 0.0257 & 0.0429 & 0.0501 & 0.0096 & 0.0339 \\
& Fine-tune & 0.0995 & 0.1534 & 0.0683 & 0.1003 & 0.1334 & 0.0923 & 0.1325 & 0.0346 & 0.0813 \\
& Joint Learning & 0.0842 & 0.1278 & 0.0553 & 0.0871 & 0.1048 & 0.0836 & 0.1040 & 0.0290 & \textbf{0.0818} \\
\cmidrule{2-11}
& Weight Averaging & 0.0886 & 0.1423 & 0.0609 & 0.0841 & 0.1017 & \uline{0.0860} & \uline{0.1372} & 0.0299 & 0.0664 \\
& Task Arithmetic & 0.0785 & 0.1268 & 0.0540 & 0.0661 & 0.0993 & 0.0772 & 0.1240 & 0.0253 & 0.0552 \\
& TIES & 0.0905 & 0.1439 & 0.0622 & 0.0881 & \textbf{0.1201} & 0.0854 & 0.1303 & \textbf{0.0324} & 0.0613 \\
& AdaMerging (Domain-wise) & 0.0689 & 0.1261 & 0.0426 & 0.0720 & 0.0683 & 0.0562 & 0.1039 & 0.0267 & 0.0556 \\
& AdaMerging (Layer-wise) & 0.0778 & 0.1301 & 0.0504 & 0.0788 & 0.0833 & 0.0766 & 0.1114 & 0.0247 & 0.0675 \\
\cmidrule{2-11}
& MergeRec (Domain-wise) & \uline{0.0913}* & \uline{0.1500}* & \uline{0.0636}* & \textbf{0.0921}* & 0.1074 & 0.0828 & 0.1363 & 0.0312 & 0.0673 \\
& MergeRec (Layer-wise) & \textbf{0.0932}* & \textbf{0.1515}* & \textbf{0.0651}* & \uline{0.0921}* & \uline{0.1102} & \textbf{0.0869}* & \textbf{0.1394}* & \uline{0.0312} & \uline{0.0696} \\

\bottomrule
\end{tabular}
\vspace{-2mm}
\end{table*}

\begin{table*}[t] \small
\caption{Performance comparison with seven baseline methods on four backbone models, \ie, RecFormer-base/large~\cite{LiWLFSSM23recformer} and BLaIR-base/large~\cite{hou2024blair}. We report absolute NDCG@10 performance. The best results, excluding the fine-tuned model, are marked in \textbf{bold}, and the second-best results are shown as \underline{underlined}. `*’ indicates the statistically significant gain of MergeRec over the best baseline model (p < 0.02 for one-tailed t-test).}

\centering
\label{tab:overall_ndcg_abs}

\begin{tabular}{c|l|>{\columncolor{gray!10}}rrrrrrrrr}
\toprule
\multicolumn{1}{c|}{\textbf{Backbone}} & 
\multicolumn{1}{c|}{\textbf{Method}} & 
\multicolumn{1}{c}{\textbf{Avg.}} & 
\multicolumn{1}{c}{Arts} & 
\multicolumn{1}{c}{Beauty} & 
\multicolumn{1}{c}{Inst.} & 
\multicolumn{1}{c}{Office} & 
\multicolumn{1}{c}{Pantry} & 
\multicolumn{1}{c}{Sci.} & 
\multicolumn{1}{c}{Sports} & 
\multicolumn{1}{c}{Toys} \\
\midrule

\multirow{9.8}{*}{RecFormer-base}
& Zero-shot & 0.0447 & 0.0717 & 0.0212 & 0.0415 & 0.0545 & 0.0381 & 0.0858 & 0.0118 & 0.0331 \\
& Fine-tune & 0.0662 & 0.1128 & 0.0344 & 0.0743 & 0.1036 & 0.0550 & 0.0924 & 0.0166 & 0.0407 \\
& Joint Learning & 0.0484 & 0.0760 & 0.0259 & 0.0566 & 0.0691 & 0.0455 & 0.0666 & 0.0116 & \textbf{0.0356} \\
\cmidrule{2-11}
& Weight Averaging & 0.0551 & 0.0935 & 0.0263 & 0.0562 & 0.0728 & 0.0508 & 0.0921 & \textbf{0.0137} & 0.0351 \\
& Task Arithmetic & 0.0557 & 0.0975 & 0.0261 & 0.0565 & 0.0791 & 0.0505 & 0.0920 & 0.0130 & 0.0307 \\
& TIES & \uline{0.0567} & \uline{0.0983} & \textbf{0.0274} & 0.0569 & \textbf{0.0842} & 0.0506 & 0.0902 & 0.0134 & 0.0326 \\
& AdaMerging (Domain-wise) & 0.0471 & 0.0851 & 0.0190 & 0.0448 & 0.0585 & 0.0391 & 0.0838 & 0.0113 & \uline{0.0354} \\
& AdaMerging (Layer-wise) & 0.0408 & 0.0687 & 0.0110 & 0.0389 & 0.0532 & 0.0353 & 0.0763 & 0.0107 & 0.0322 \\
\cmidrule{2-11}
& MergeRec (Domain-wise) & \textbf{0.0570}* & \textbf{0.0985} & 0.0263 & \textbf{0.0588}* & \uline{0.0796} & \textbf{0.0520}* & \uline{0.0929}* & 0.0134 & 0.0346 \\
& MergeRec (Layer-wise) & 0.0565 & 0.0972 & \uline{0.0266} & \uline{0.0578}* & 0.0781 & \uline{0.0516}* & \textbf{0.0933}* & \uline{0.0136} & 0.0341 \\
\midrule

\multirow{9.8}{*}{RecFormer-large}
& Zero-shot & 0.0357 & 0.0720 & 0.0174 & 0.0260 & 0.0317 & 0.0235 & 0.0788 & 0.0096 & 0.0266 \\
& Fine-tune & 0.0667 & 0.1107 & 0.0346 & 0.0764 & 0.1015 & 0.0574 & 0.0954 & 0.0157 & 0.0419 \\
& Joint Learning & 0.0505 & 0.0798 & 0.0261 & 0.0508 & 0.0677 & 0.0499 & 0.0836 & 0.0134 & 0.0329 \\
\cmidrule{2-11}
& Weight Averaging & 0.0573 & 0.0956 & 0.0283 & 0.0584 & 0.0757 & \uline{0.0543} & 0.0941 & \textbf{0.0143} & 0.0372 \\
& Task Arithmetic & 0.0572 & 0.1036 & 0.0277 & 0.0593 & 0.0786 & 0.0526 & 0.0902 & 0.0128 & 0.0326 \\
& TIES & 0.0585 & \uline{0.1042} & 0.0283 & \uline{0.0616} & \textbf{0.0833} & 0.0530 & 0.0920 & 0.0131 & 0.0322 \\
& AdaMerging (Domain-wise) & 0.0449 & 0.0853 & 0.0197 & 0.0473 & 0.0578 & 0.0319 & 0.0791 & 0.0088 & 0.0292 \\
& AdaMerging (Layer-wise) & 0.0433 & 0.0786 & 0.0179 & 0.0410 & 0.0555 & 0.0357 & 0.0776 & 0.0110 & 0.0293 \\
\cmidrule{2-11}
& MergeRec (Domain-wise) & \textbf{0.0596}* & \textbf{0.1066}* & \uline{0.0293}* & \textbf{0.0617} & \uline{0.0794} & 0.0540 & \uline{0.0944} & 0.0138 & \uline{0.0379}* \\
& MergeRec (Layer-wise) & \uline{0.0591}* & 0.1034 & \textbf{0.0295}* & 0.0598 & 0.0788 & \textbf{0.0544} & \textbf{0.0950} & \uline{0.0141} & \textbf{0.0380}* \\
\midrule

\multirow{9.8}{*}{BLaIR-base}
& Zero-shot & 0.0218 & 0.0368 & 0.0103 & 0.0178 & 0.0216 & 0.0238 & 0.0426 & 0.0052 & 0.0165 \\
& Fine-tune & 0.0660 & 0.1119 & 0.0309 & 0.0742 & 0.1038 & 0.0582 & 0.0931 & 0.0153 & 0.0404 \\
& Joint Learning & \uline{0.0518} & 0.0895 & \textbf{0.0301} & \textbf{0.0588} & 0.0702 & \textbf{0.0512} & 0.0633 & \textbf{0.0141} & \textbf{0.0372} \\
\cmidrule{2-11}
& Weight Averaging & 0.0516 & 0.0879 & \uline{0.0265} & 0.0497 & 0.0672 & \uline{0.0498} & \textbf{0.0881} & 0.0126 & 0.0306 \\
& Task Arithmetic & 0.0365 & 0.0561 & 0.0160 & 0.0412 & 0.0543 & 0.0342 & 0.0620 & 0.0075 & 0.0209 \\
& TIES & 0.0493 & 0.0808 & 0.0226 & 0.0487 & \textbf{0.0816} & 0.0418 & 0.0768 & \uline{0.0126} & 0.0296 \\
& AdaMerging (Domain-wise) & 0.0332 & 0.0577 & 0.0152 & 0.0308 & 0.0367 & 0.0251 & 0.0615 & 0.0105 & 0.0279 \\
& AdaMerging (Layer-wise) & 0.0382 & 0.0637 & 0.0176 & 0.0398 & 0.0465 & 0.0288 & 0.0723 & 0.0106 & 0.0263 \\
\cmidrule{2-11}
& MergeRec (Domain-wise) & 0.0517 & \textbf{0.0910}* & 0.0253 & 0.0487 & \uline{0.0732} & 0.0455 & 0.0858 & 0.0122 & \uline{0.0316} \\
& MergeRec (Layer-wise) & \textbf{0.0522} & \uline{0.0909}* & 0.0261 & \uline{0.0497} & 0.0729 & 0.0477 & \uline{0.0869} & 0.0125 & 0.0313 \\
\midrule

\multirow{9.8}{*}{BLaIR-large}
& Zero-shot & 0.0172 & 0.0301 & 0.0091 & 0.0134 & 0.0141 & 0.0237 & 0.0278 & 0.0045 & 0.0150 \\
& Fine-tune & 0.0642 & 0.1093 & 0.0328 & 0.0743 & 0.0991 & 0.0565 & 0.0865 & 0.0160 & 0.0388 \\
& Joint Learning & 0.0524 & 0.0886 & 0.0265 & \textbf{0.0635} & \uline{0.0752} & \textbf{0.0521} & 0.0621 & 0.0136 & \textbf{0.0376} \\
\cmidrule{2-11}
& Weight Averaging & 0.0523 & 0.0910 & 0.0274 & 0.0533 & 0.0672 & 0.0513 & 0.0851 & 0.0135 & 0.0296 \\
& Task Arithmetic & 0.0468 & 0.0802 & 0.0254 & 0.0425 & 0.0656 & 0.0462 & 0.0775 & 0.0121 & 0.0253 \\
& TIES & \uline{0.0550} & 0.0937 & \uline{0.0289} & 0.0567 & \textbf{0.0828} & 0.0508 & 0.0847 & \textbf{0.0145} & 0.0279 \\
& AdaMerging (Domain-wise) & 0.0386 & 0.0748 & 0.0191 & 0.0428 & 0.0421 & 0.0317 & 0.0623 & 0.0121 & 0.0242 \\
& AdaMerging (Layer-wise) & 0.0454 & 0.0810 & 0.0227 & 0.0497 & 0.0541 & 0.0456 & 0.0684 & 0.0113 & 0.0300 \\
\cmidrule{2-11}
& MergeRec (Domain-wise) & 0.0542 & \uline{0.0948}* & 0.0281 & 0.0584 & 0.0717 & 0.0494 & \uline{0.0872}* & 0.0138 & 0.0300 \\
& MergeRec (Layer-wise) & \textbf{0.0556}* & \textbf{0.0958}* & \textbf{0.0292}* & \uline{0.0588} & 0.0743 & \uline{0.0517} & \textbf{0.0901}* & \uline{0.0140} & \uline{0.0310} \\

\bottomrule
\end{tabular}
\vspace{-2mm}
\end{table*}

\end{document}